\documentclass[12pt,a4paper]{article}
\usepackage[left=2.5cm,right=2.5cm,top=3.5cm,bottom=3.5cm,a4paper]{geometry}
\pdfoutput=1 

\usepackage[T1]{fontenc} 
\usepackage[utf8]{inputenc}

\usepackage{xspace}

\newcommand{\Dz}{{\ensuremath{D^0}}\xspace}
\newcommand{\Dp}{{\ensuremath{D^+}}\xspace}
\newcommand{\Dstar}{{\ensuremath{D^{*+}}}\xspace}
\newcommand{\Dstarbar}{{\ensuremath{\bar{D}^{*-}}}\xspace}
\newcommand{\B}{{\ensuremath{B}}\xspace}
\newcommand{\Bbar}{{\ensuremath{\bar{B}}}\xspace}
\newcommand{\nul}{{\ensuremath{\nu_{l}}}\xspace}

\newcommand{\BtoDstarlnu}{{\ensuremath{\Bbar\to\Dstar l^-\bar{\nu}_{l}}}\xspace}
\newcommand{\BtoDstarlnuCP}{{\ensuremath{\B\to \Dstarbar l^+\nu_{l}}}\xspace}
\newcommand{\BtoDstarlnutod}{{\ensuremath{\Bbar\to\Dstar(\to\Dz\pi^+) l^-\bar{\nu}_{l}}}\xspace}
\newcommand{\BtoDstarmunu}{{\ensuremath{\Bbar\to\Dstar\mu^-\bar{\nu}_{\mu}}}\xspace}
\newcommand{\BtoDstartaunu}{{\ensuremath{\Bbar\to\Dstar\tau^-\bar{\nu}_{\tau}}}\xspace}
\newcommand{\BtoDstartaunutopi}{{\ensuremath{\Bbar\to\Dstar\tau^-(\to \pi^+\pi^-\pi^-\nu_{\tau})\bar{\nu}_{\tau}}}\xspace}
\newcommand{\BtoDstartaunutomu}{{\ensuremath{\Bbar\to\Dstar\tau^-(\to \mu^-\bar{\nu}_{\mu}\nu_{\tau})\bar{\nu}_{\tau}}}\xspace}

\newcommand{\thetaD}{{\ensuremath{\theta_{\Dz}}}\xspace}
\newcommand{\thetal}{{\ensuremath{\theta_{l}}}\xspace}
\newcommand{\costhetaD}{{\ensuremath{\cos\thetaD}}\xspace}
\newcommand{\costhetal}{\ensuremath{\cos\thetal}\xspace}
\newcommand{\cosqthetaD}{{\ensuremath{\cos^2\thetaD}}\xspace}

\newcommand{\fl}{{\ensuremath{\mathcal{F}_L}}\xspace}
\newcommand{\ft}{{\ensuremath{\mathcal{F}_T}}\xspace}
\newcommand{\At}{{\ensuremath{\mathcal{A}^{(1)}_T}}\xspace}
\newcommand{\Atbar}{{\ensuremath{\bar{\mathcal{A}}^{(1)}_T}}\xspace}
\newcommand{\Acp}{{\ensuremath{\mathcal{A}^{(1)}_{\CP}}}\xspace}
\newcommand{\Ac}{{\ensuremath{\mathcal{A}^{(1)}_C}}\xspace}
\newcommand{\Att}{{\ensuremath{\mathcal{A}^{(2)}_T}}\xspace}
\newcommand{\Atz}{{\ensuremath{\mathcal{A}^{(0)}_T}}\xspace}
\newcommand{\Acz}{{\ensuremath{\mathcal{A}^{(0)}_C}}\xspace}
\newcommand{\Attbar}{{\ensuremath{\bar{\mathcal{A}}^{(2)}_T}}\xspace}
\newcommand{\Acpp}{{\ensuremath{\mathcal{A}^{(2)}_{\CP}}}\xspace}
\newcommand{\Acc}{{\ensuremath{\mathcal{A}^{(2)}_C}}\xspace}
\newcommand{\TPA}{{\ensuremath{\mathrm{TPA}^{(1)}}}\xspace}
\newcommand{\TPAz}{{\ensuremath{\mathrm{TPA}^{(0)}}}\xspace}
\newcommand{\TPAA}{{\ensuremath{\mathrm{TPA}^{(2)}}}\xspace}
\newcommand{\RDstar}{{\ensuremath{R(D^{*})}}\xspace}

\newcommand{\tev}{\ensuremath{\mathrm{\,Te\kern -0.1em V}}\xspace}
\newcommand{\gev}{\ensuremath{\mathrm{\,Ge\kern -0.1em V}}\xspace}
\newcommand{\mev}{\ensuremath{\mathrm{\,Me\kern -0.1em V}}\xspace}
\newcommand{\kev}{\ensuremath{\mathrm{\,ke\kern -0.1em V}}\xspace}
\newcommand{\mm}{\ensuremath{\mathrm{ \,mm}}\xspace}
\newcommand{\mum}{\ensuremath{{\,\mu\mathrm{m}}}\xspace}

\newcommand{\CP}{{\ensuremath{C\!P}}\xspace}

\newcommand{\splot}{\mbox{{\em s}Plot}\xspace}
\usepackage{amsmath,amssymb,xspace,hyperref,graphicx,color,ifpdf,ifthen,mciteplus}
\usepackage{booktabs,multirow,cite}
\usepackage{tikz}
\graphicspath{{figs/}} 

\title{\boldmath \textbf{Angular and \CP-violation analyses of \BtoDstarlnu decays at hadron collider experiments}}

 \author{\textbf{Daniele Marangotto}\footnote{E-mail: \texttt{daniele.marangotto@unimi.it}}\\[2ex]Universit\`a degli studi di Milano and INFN Milano}

\newboolean{articletitles}
\setboolean{articletitles}{true} 

\newboolean{inbibliography}
\setboolean{inbibliography}{false} 

\begin{document} 
\maketitle

\section*{Abstract}
The \BtoDstarlnu branching fractions ratio between muon and tau lepton decay modes \RDstar has shown intriguing discrepancies between the Standard Model prediction and measurements performed at BaBar, Belle and LHCb experiments, a possible sign of beyond the Standard Model physics. Theoretical studies show how observables related to the \BtoDstarlnu differential decay distribution can be used to further constrain New Physics contributions, but their experimental measurements is lacking to date.
This article proposes the measurement of \BtoDstarlnu angular and \CP-violating observables at hadron collider experiments, by exploiting approximate reconstruction algorithms using information from detectable final-state particles only. The resolution on the phase space variables is studied using \BtoDstarlnu decays simulated in a forward detector geometry like LHCb. A method to correct the observable values for the reconstruction inaccuracies based on detector simulation is successfully tested on simulated data and the decrease in precision with respect to a perfect reconstruction is evaluated. The \Dstar longitudinal polarization fraction and the \Att \CP-violating observable can be measured losing a factor 2 and 5 in precision, respectively. The extraction of angular distributions from the template fit selecting \BtoDstarlnu decays and associated systematic uncertainties are also discussed.

\clearpage

\section{Introduction}
\label{sec:intro}
Semileptonic \BtoDstarlnu decays, in which $l^-$ stands for one of the three charged leptons, have shown intriguing discrepancies between the Standard Model predicted ratio of branching fractions between muon and tau lepton decay modes~\cite{Fajfer:2012vx}, indicated as \RDstar, and the measured values at BaBar~\cite{Lees:2013uzd}, Belle~\cite{Huschle:2015rga,Sato:2016svk,Hirose:2016wfn} and LHCb~\cite{Aaij:2015yra,Aaij:2017deq} experiments. This contrast could be a sign of New Physics contributions violating the Standard Model universality of leptonic interactions.

The measurement of observables related to the \BtoDstarlnu differential decay rate, other than \RDstar, can shed new light on the observed anomalies, allowing to put complementary constraints on possible New Physics sources~\cite{Fajfer:2012vx,Tanaka:2012nw,Duraisamy:2013kcw,Bhattacharya:2015ida,Alok:2016qyh,Colangelo:2018cnj,Bhattacharya:2018kig}. However, the only measurement of these observables available to date is a preliminary result for the \Dstar longitudinal polarization fraction in \BtoDstartaunu decays by the Belle experiment~\cite{Adamczyk:2019wyt}
\begin{equation}
\fl =  0.60 \pm 0.08(stat.) \pm 0.04(syst.),
\end{equation}
which is consistent at 1.4$\sigma$ with the Standard Model prediction $\fl = 0.46 \pm 0.04$~\cite{Alok:2016qyh,Bhattacharya:2018kig}.

Angular analyses of \BtoDstarlnu decays are challenging because final-state neutrinos can not be reconstructed, implying that the \Bbar meson rest frame is not precisely determined from the detectable part of the decay. This problem can be mitigated at $e^+e^-$ \B-factories, where the momentum of the \Bbar meson can be determined from the known center-of-mass energy of the $e^+e^-$ collision and the complete reconstruction of the decay of the other \B meson produced in the interaction. On the contrary, at hadronic colliders the \Bbar meson momentum is not constrained by the production mechanism since the center-of-mass energy of the parton-parton collision is unknown.

This article considers the possibility to measure the angular variable distributions of \BtoDstarlnu decays by exploiting reconstruction algorithms estimating the \Bbar meson rest frame only from information related to the detectable final-state particles, a situation of particular interest for hadron collider experiments like LHCb. The attainable precision on the phase space variables is studied by means of a simulation study set for a forward detector geometry which is detailed in section~\ref{sec:reconstruction}. It is shown that observables related to the cosine of the polar angle of the \Dz meson in the \Dstar helicity frame, \costhetaD, and the azimuthal angle between the ($\Dz\pi^+$) and ($l^-\nu$) decay planes, $\chi$, are suitable to be measured in the considered set-up. It is shown that \costhetaD and $\chi$ distributions can be extracted using the \splot statistical technique~\cite{Pivk:2004ty} from the template fit selecting \BtoDstarlnu decays from background events.

The fully differential \BtoDstarlnu decay distribution is reviewed in section~\ref{sec:decay_distribution} and the observables associated to the aforementioned phase space distributions introduced. These are the \Dstar longitudinal polarization, the \CP-conserving and \CP-violating observables related to the $\chi$ angle distributions. The latter are especially interesting being a null test for the Standard Model, since \CP-violation in Cabibbo-favoured $b \to c$ quark transition is strongly suppressed by the Cabibbo-Kobayashi-Maskawa mechanism.

In section~\ref{sec:measurement}, a method to measure the considered observables while correcting the effect of reconstruction inaccuracies is presented and tested on simulated \BtoDstarlnu decays. The decrease in precision due to the use of the reconstruction algorithms is evaluated with respect to ideal measurements in which the phase space distributions are perfectly reconstructed. A discussion on the possible systematic uncertainties associated to the proposed measurements is reported in section~\ref{sec:systematic_uncertainties}. The conclusions of the study are summarized in section~\ref{sec:conclusions}.

\section{The \BtoDstarlnu decay reconstruction}
\label{sec:reconstruction}
\subsection{Simulation configuration}
The capability of reconstructing the \BtoDstarlnutod decay distribution using approximate reconstruction algorithms is studied on simulated semileptonic decays in a detector configuration analogous to the LHCb experiment~\cite{Alves:2008zz}.

Three decay chains are considered: \BtoDstarmunu, \BtoDstartaunutopi and  \BtoDstartaunutomu, along with their charge-conjugated decays. The flavour of the \Bbar meson is determined by the charge of the detectable part of the lepton decay or by that of the pion produced in the $\Dstar\to\Dz\pi^+$ decay. The production of \Bbar mesons from proton-proton collisions at a center-of-mass energy $\sqrt{s}=13\tev$ are simulated using PYTHIA 8.1~\cite{Sjostrand:2006za,Sjostrand:2007gs}, their decay to the different final states are simulated by the EVTGEN package~\cite{Lange:2001uf}. Stable particles are required to be within the nominal LHCb pseudorapidity acceptance $2<\eta<5$, while charged particle momentum cuts $p_T>250 \mev$ and $p>5 \gev$ roughly reproducing the LHCb kinematic acceptance (estimated from~\cite{Alves:2008zz}) have been tried but showed no significant effect on the subsequent studies. A minimum \Bbar meson flight distance of 3\mm simulates the effect of a displaced vertex trigger requirement. The production and decay vertex positions of the \Bbar meson have been smeared from their generated values according to Gaussian distributions reproducing the performance of the LHCb VELO detector~\cite{LHCb:2001aa,LHCbVELOGroup:2014uea}: for production vertexes the Gaussian widths are 13\mum and 70\mum in the transverse and longitudinal directions, respectively, with respect to the beam; for decay vertexes they are 20\mum and 200\mum. For \BtoDstartaunutopi decays, a minimum tau lepton flight distance of 1\mm is applied as background rejection cut.

The ROOT package~\cite{Brun:1997pa} is employed for data handling and graphics.

\subsection{\Bbar rest frame approximate reconstruction algorithms}
The \Bbar rest frame reconstruction benefits from the knowledge of the flight direction from its production and decay vertexes, the latter determined by the $\Dstar(\to D^0(\to K^-\pi^+) \pi^+)$ track combination. Two strategies are considered in this study.

For decays in which a single neutrino is missing, the available information about the decay (the momentum of the detectable part of the decay, the \Bbar meson flight direction, the \Bbar and neutrino masses) determines the \Bbar momentum up to a two-fold ambiguity~\cite{Dambach:2006ha}. The two solutions correspond to the forward or backward orientation of the neutrino in the \Bbar rest frame with respect to the \Bbar flight direction. If the neutrino is orthogonal to the \Bbar flight direction a unique, degenerate solution is found.
This algorithm will be referred to as "full reconstruction".

A different \Bbar momentum approximation can be made assuming that the proper velocity along the beam axis, $\gamma\beta_z$, of the detectable part of the decay is equal to that of the \Bbar meson~\cite{Aaij:2015yra}. The magnitude of the \Bbar momentum in terms of the visible decay system $V$ and the angle $\theta$ between flight direction and beam axis is set as
\begin{equation}
|p(\Bbar)| = p_z(V)\frac{m(\Bbar)}{m(V)}\sqrt{1+\tan^2\theta}.
\end{equation}
This approach will be referred to as "equal velocity" algorithm and it is applicable also to decays with two or more invisible particles, in which the invariant mass of the unmeasured part of the decay is unknown.

\subsection{Resolutions on the \BtoDstarlnu phase space variables}
\label{sec:resolutions}
\begin{figure}
\centering
\begin{tikzpicture}[scale=2.2]
\draw (-1,0) -- (1,0);
\draw (0,0) circle [radius=0.05];
\node [below right] at (0,0) {\Bbar};
\draw [dashed] (1,0) -- (2,0);
\draw [dashed] (-1,0) -- (-2,0);
\node [below right] at (2,0) {$\hat{z}$};
\draw [->] (1,0) -- (1.4,0.7);
\draw [->] (1,0) -- (0.6,-0.7);
\node [below right] at (1,0) {\Dstar};
\node [right] at (1.4,0.7) {\Dz};
\node [right] at (0.6,-0.7) {$\pi^+$};
\node [above right] at (1.3,0.1) {\thetaD};
\draw (1.3,0) arc [radius=0.3,start angle=0,end angle=60.255];
\draw (1.33,0) arc [radius=0.33,start angle=0,end angle=60.255];
\draw [->] (-1,0) -- (-1.2,0.78);
\draw [->] (-1,0) -- (-0.8,-0.78);
\draw (-1.3,0) arc [radius=0.3,start angle=180,end angle=104.381];
\draw (-1.33,0) arc [radius=0.33,start angle=180,end angle=104.381];
\node [left] at (-1.2,0.78) {$l^{-}$};
\node [left] at (-0.8,-0.78) {$\bar{\nu}_l$};
\node [above left] at (-1.3,0.1) {\thetal};
\draw (-0.5,-0.875) -- (0.5,0.875);
\draw (1.5,-0.875) -- (2.5,0.875);
\draw (0.5,0.875) -- (2.5,0.875);
\draw (-0.5,-0.875) -- (1.5,-0.875);
\draw (-0.25,-0.975) -- (0.25,0.975);
\draw (-2.25,-0.975) -- (-1.75,0.975);
\draw (-0.25,-0.975) -- (-2.25,-0.975);
\draw (0.25,0.975) -- (-1.75,0.975);
\draw (0.2954,0.517) arc [radius=0.6,start angle=60.255,end angle=75.619];
\draw (0.31017,0.54285) arc [radius=0.63,start angle=60.255,end angle=75.619];
\node [above] at (0.26017,0.58285) {$\chi$};
\end{tikzpicture}
\caption{Definition of the \BtoDstarlnu phase space variables.\label{fig:phase_space}}
\end{figure}
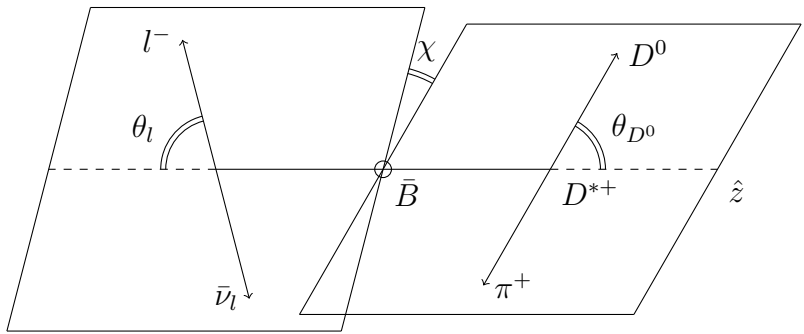

The \BtoDstarlnutod decay is characterized by four degrees of freedom.\footnote{The $\Dz\pi^+$ invariant mass is considered fixed given the very small \Dstar width $83.4\pm 1.8 \kev$~\cite{Tanabashi:2018oca}.} Its phase space can be described by the following four kinematic variables: the invariant mass of the $l^-\bar{\nu}_{l}$ system $q^2$, the cosine of the polar angle of the \Dz meson in the \Dstar helicity frame \costhetaD, the cosine of the polar angle of the lepton in the $l^-\bar{\nu}_{l}$ system helicity frame \costhetal and the azimuthal angle between the ($\Dz\pi^+$) and ($l^-\bar{\nu}_{l}$) decay planes $\chi$, see figure~\ref{fig:phase_space}. In \Dstar and $l^-\bar{\nu}_{l}$ helicity frames, the $z$ axis is defined by the direction of the \Dstar and $l^-\bar{\nu}_{l}$ momenta in the \Bbar rest frame, respectively.

The attainable precision on the four phase space variables is studied computing the resolution defined as the difference between the values measured using the reconstruction algorithms and the true values of the simulated events. Differences of dimensional quantities are divided by the true values.

\begin{figure}
\centering
\includegraphics[scale=0.35]{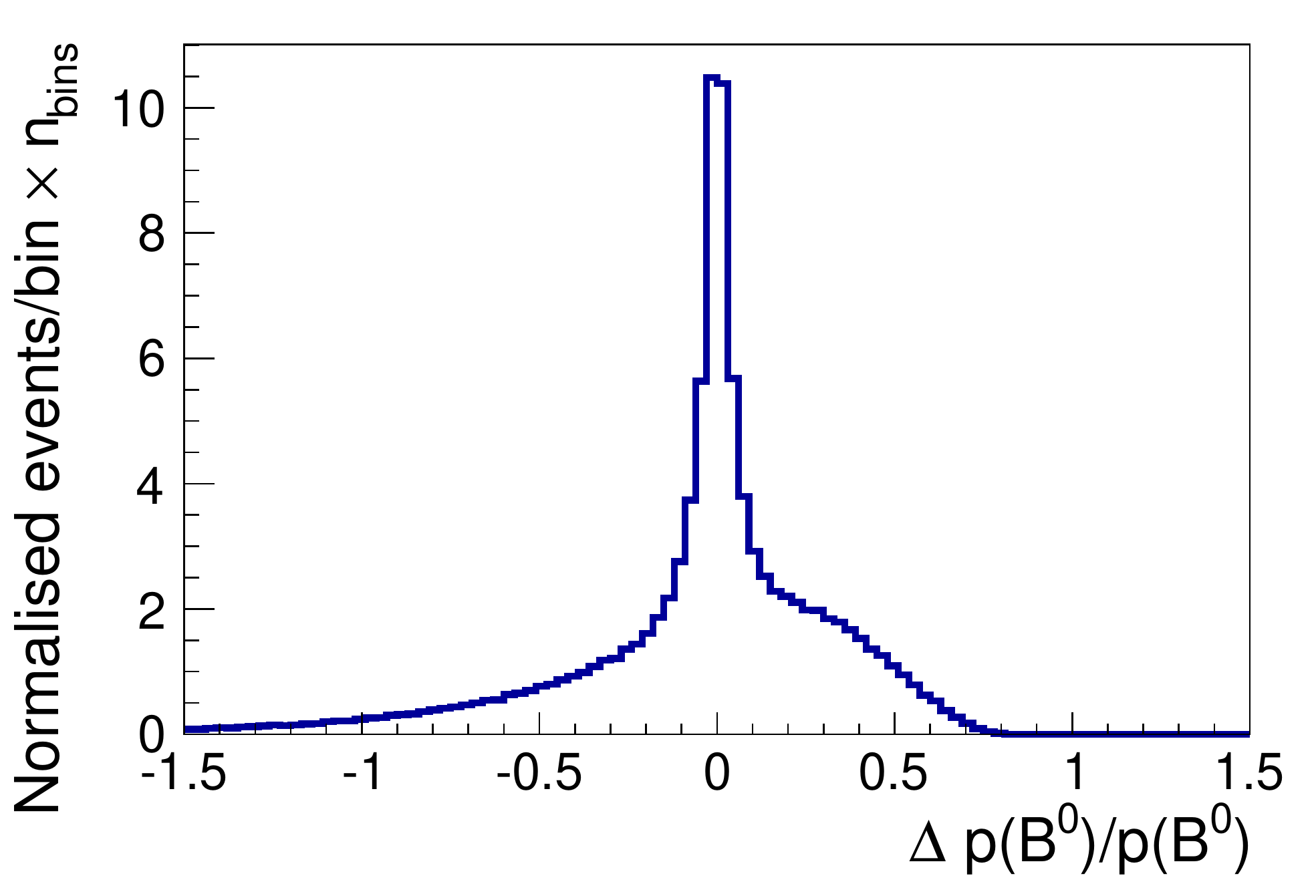}
\includegraphics[scale=0.35]{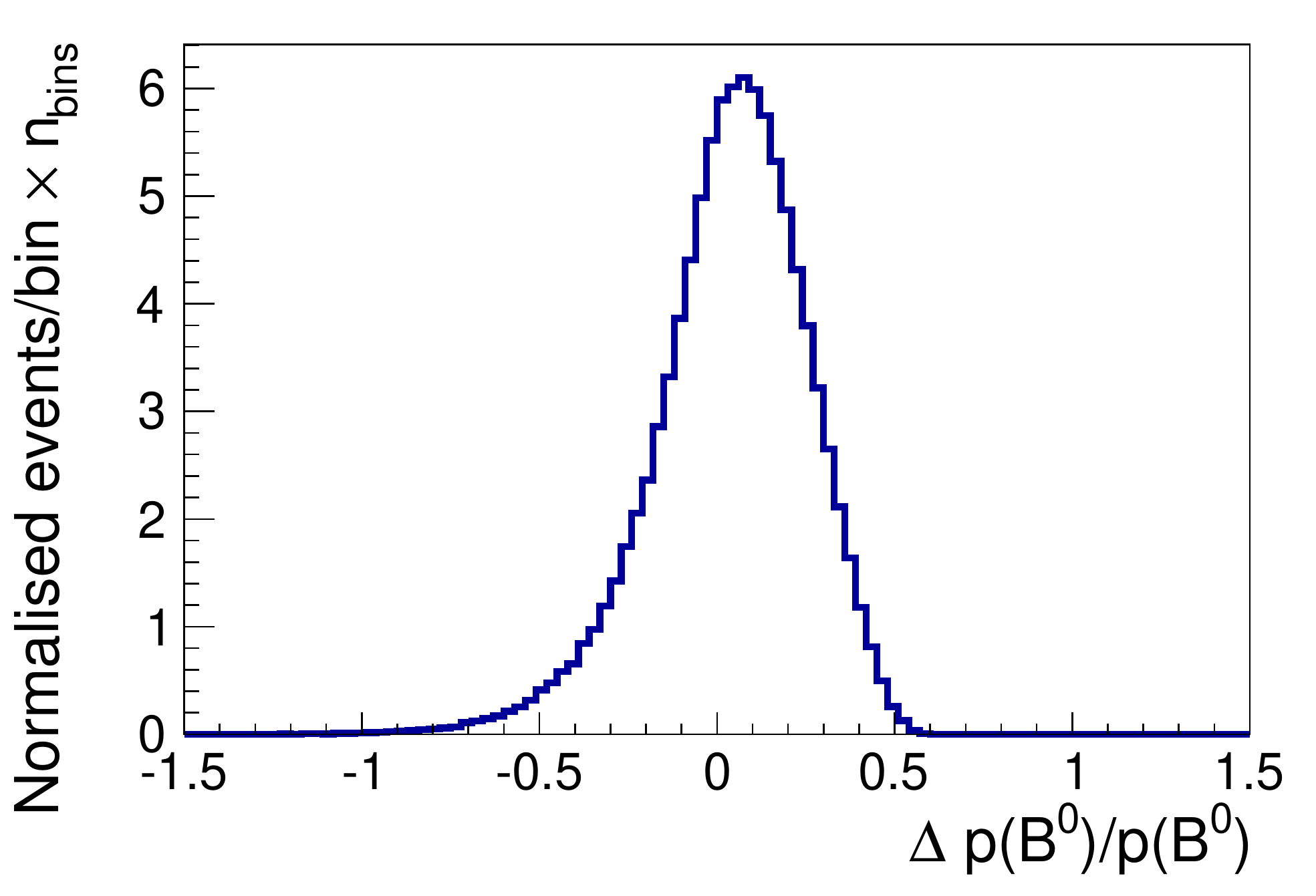}
\caption{Relative resolution on the \Bbar momentum magnitude for the \BtoDstarmunu decay, obtained using (left) full reconstruction and (right) equal velocity algorithms.\label{fig:B2Dstmunu_B_P_res}}
\end{figure}

\begin{figure}
\centering
\includegraphics[scale=0.7]{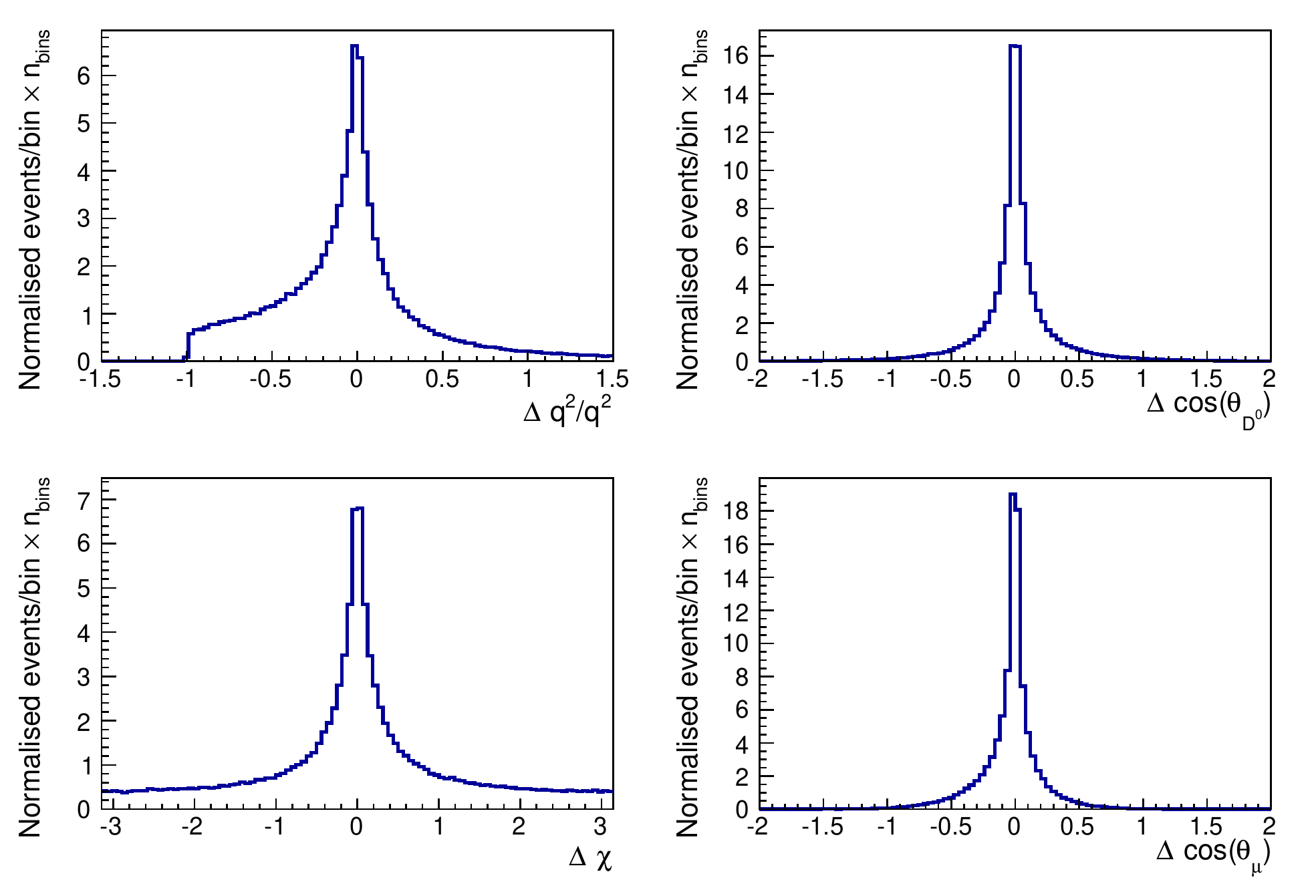}
\caption{Resolution on the phase space variables for \BtoDstarmunu decays, obtained with full reconstruction algorithm. \label{fig:B2Dstmunu_distributions_res}}
\end{figure}

\begin{figure}
\centering
\includegraphics[scale=0.7]{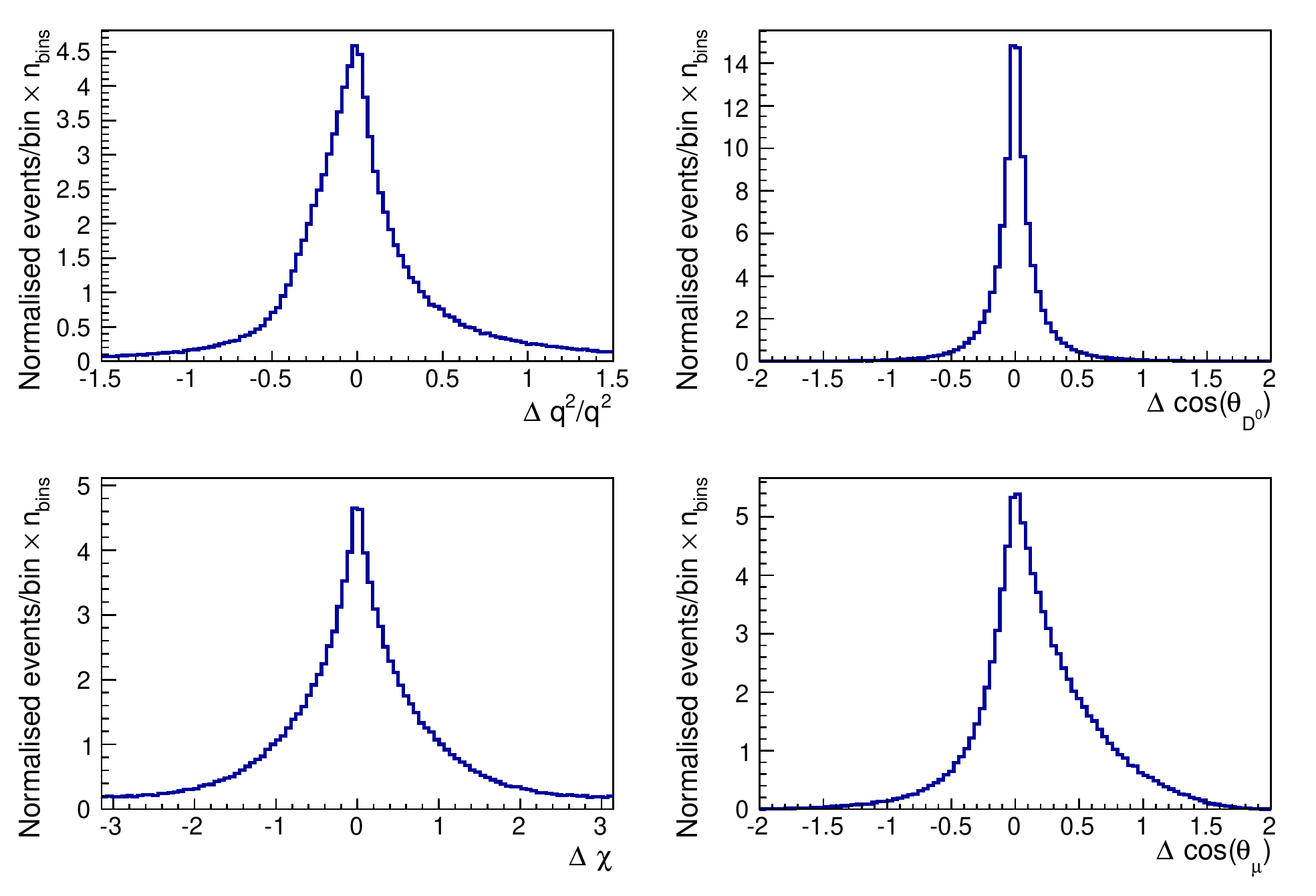}
\caption{Resolution on the phase space variables for \BtoDstarmunu decays, obtained with equal velocity algorithm. \label{fig:B2Dstmunu_distributions_res_boost}}
\end{figure}

The \Bbar rest frame reconstruction for \BtoDstarmunu decays is achieved exploiting the full reconstruction algorithm. If a couple of solutions are found, one of the two is selected by random choice, while apparently unphysical configurations, due to experimental uncertainties, in which no \Bbar momentum solution is available are discarded from the following study, these constituting the 32.7\% of the simulated events. Regression techniques based on \Bbar meson flight direction and magnitude to improve the solution decision~\cite{Ciezarek:2016lqu} have been tried but showed limited improvement. The relative resolution on the \Bbar momentum magnitude, obtained with the two reconstruction algorithms, is shown in figure~\ref{fig:B2Dstmunu_B_P_res}.
The full reconstruction \Bbar momentum resolution features a narrow, symmetric distribution peaked at zero, corresponding to events in which the momentum solution corresponding to the true orientation of the neutrino (forward or backward) was chosen, and a broader, asymmetric shape associated to events in which the momentum solution corresponding to the wrong neutrino orientation was assigned. The equal velocity reconstruction presents a more regular but wider distribution.
The phase space variables describing the semileptonic decay are computed in the \Bbar rest frame resulting from the estimated \Bbar momentum. Their resolutions are reported in figure~\ref{fig:B2Dstmunu_distributions_res}: the \costhetaD and $\chi$ feature symmetric and unbiased distributions, the \costhetal distribution is slightly asymmetric but almost unbiased and the relative $q^2$ even if asymmetric peaks at zero.
Phase space variable resolutions obtained with the equal velocity algorithm are reported in figure~\ref{fig:B2Dstmunu_distributions_res_boost}. Their distributions are wider than those resulting from the full reconstruction algorithm, since less information on the decay is employed.

\begin{figure}
\centering
\includegraphics[scale=0.35]{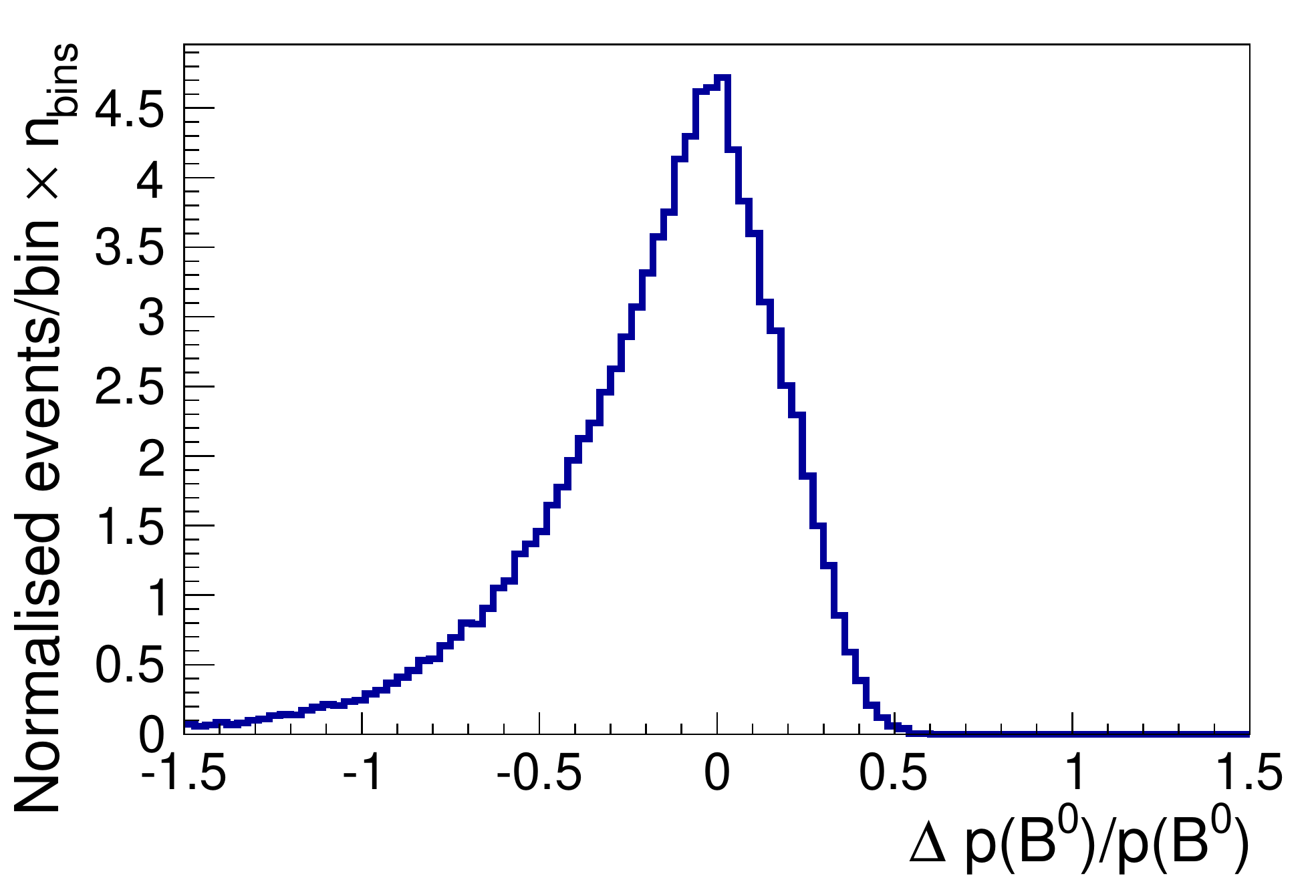}
\includegraphics[scale=0.35]{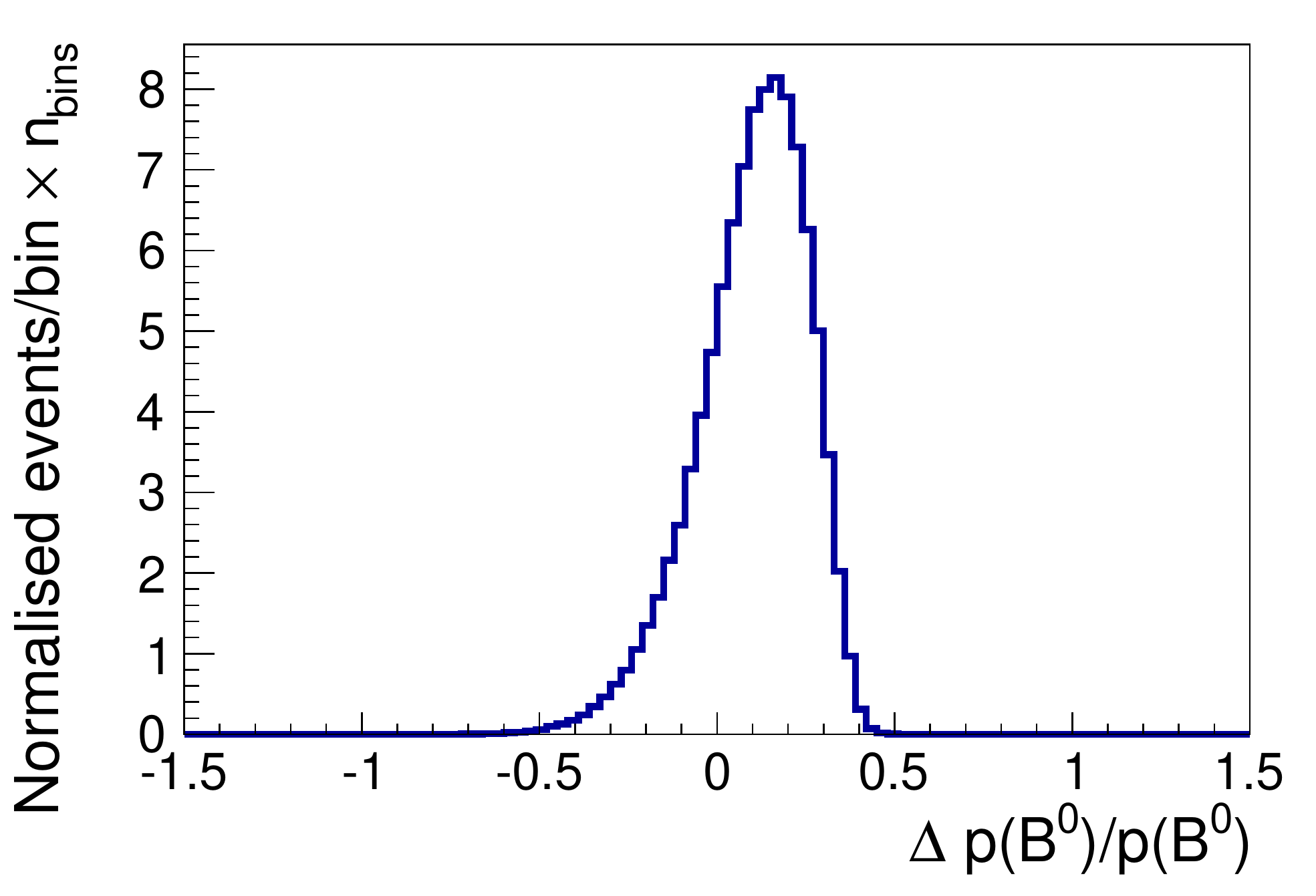}
\caption{Relative resolution on the \Bbar momentum magnitude for \BtoDstartaunutopi decays, obtained using (left) full reconstruction and (right) equal velocity algorithms.\label{fig:B2Dsttaunupi_B_P_res}}
\end{figure}

\begin{figure}
\centering
\includegraphics[scale=0.7]{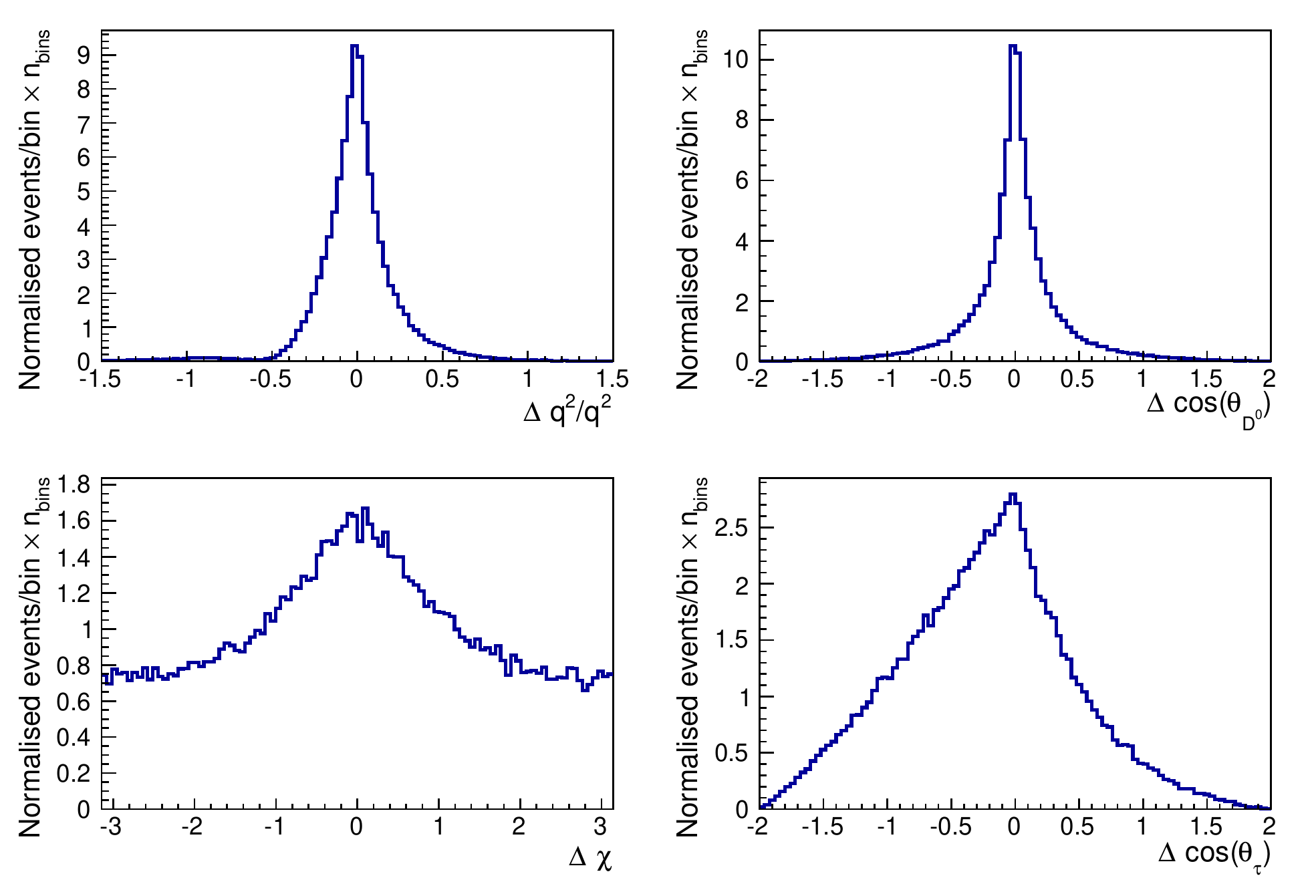}
\caption{Resolution on the phase space variables for \BtoDstartaunutopi decays, obtained with full reconstruction algorithm. \label{fig:B2Dsttaunupi_distributions_res}}
\end{figure}

\begin{figure}
\centering
\includegraphics[scale=0.7]{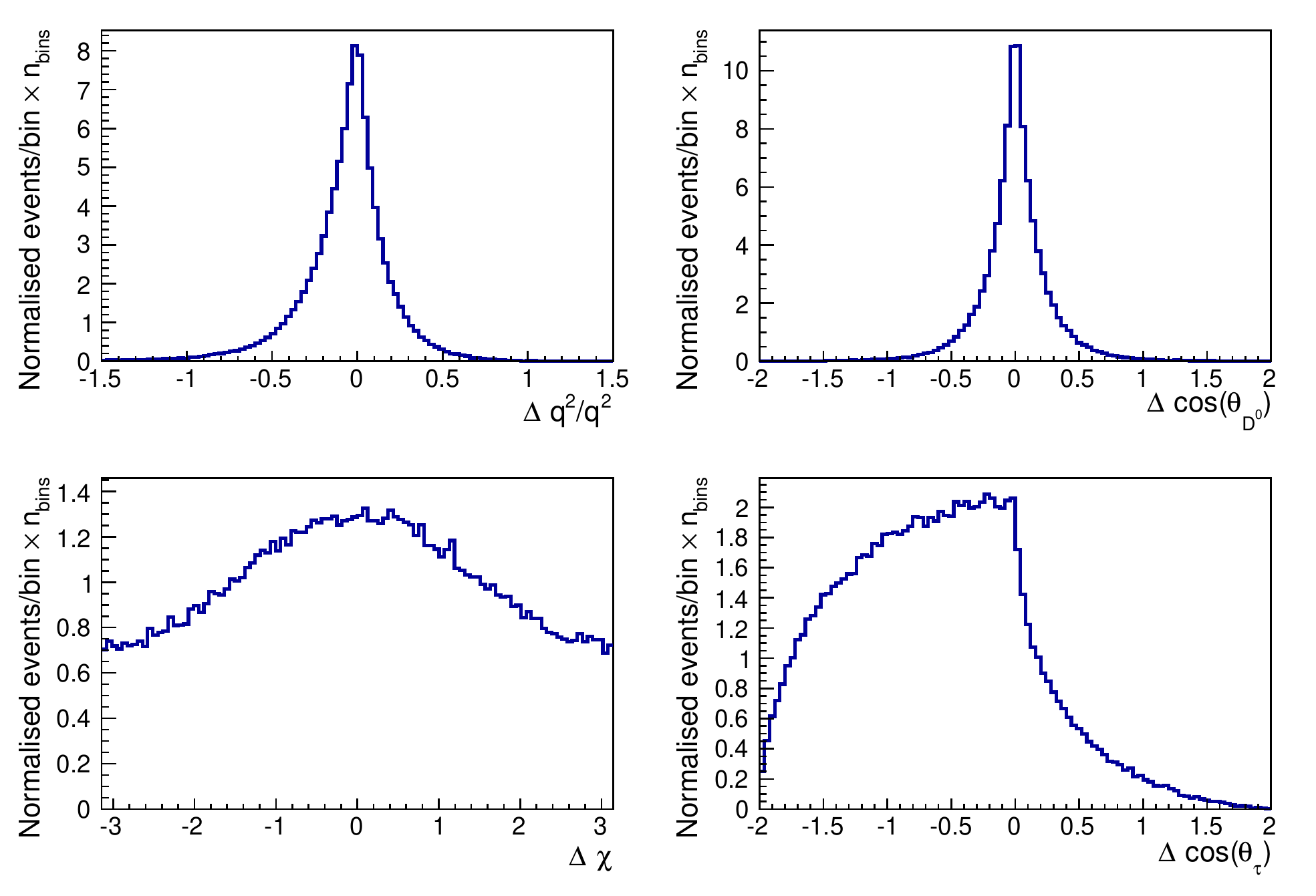}
\caption{Resolution on the phase space variables for \BtoDstartaunutopi decays, obtained with equal velocity algorithm. \label{fig:B2Dsttaunupi_distributions_res_boost}}
\end{figure}

For \BtoDstartaunutopi decays, in which the $\Dstar$ and $\pi^+\pi^-\pi^-$ vertexes determine the flight direction of the tau lepton, the full reconstruction algorithm is applied sequentially to the tau lepton and \Bbar meson decays. First, the $\tau$ momentum is estimated from the visible $3\pi$ system: if there are two $\tau$ momentum solutions one is chosen randomly. If no solutions are available, the momentum corresponding to the degenerate solution is assigned. Then, the \Bbar momentum is calculated from the $\Dstar\tau^-$ system using the estimated $\tau$ momentum: if there are two \Bbar momentum solutions one is chosen randomly. If no solutions are available then the other, if any, $\tau$ momentum solution is tried, and the event discarded only if the \Bbar momentum reconstruction is still impossible. This algorithm tries to retain the maximum information on the decay, however, it rejects 57.7\% of the events. The estimated $\tau$ momentum is then used for computing $\chi$ and \costhetal variables.
The relative resolution on the \Bbar momentum magnitude is shown in figure~\ref{fig:B2Dsttaunupi_B_P_res} along with that obtained using the equal velocity algorithm, the latter being the narrower one. Phase space variables resolutions for full reconstruction algorithm are reported in figure~\ref{fig:B2Dsttaunupi_distributions_res}, which are to be compared to those obtained applying the equal velocity algorithm, see figure~\ref{fig:B2Dsttaunupi_distributions_res_boost}. Comparing to the muon channel, the \costhetaD distributions are moderately wider, while $\chi$ and \costhetal resolutions are significantly broader, since they directly depend on the leptonic part of the decay. The $\chi$ distributions are however still unbiased, while the \costhetal ones are asymmetric and biased, especially for the equal velocity algorithm. Comparing the two algorithms, the \costhetaD distributions are basically equal, while the $\chi$ resolution is better for the full reconstruction one.


\begin{figure}
\centering
\includegraphics[scale=0.35]{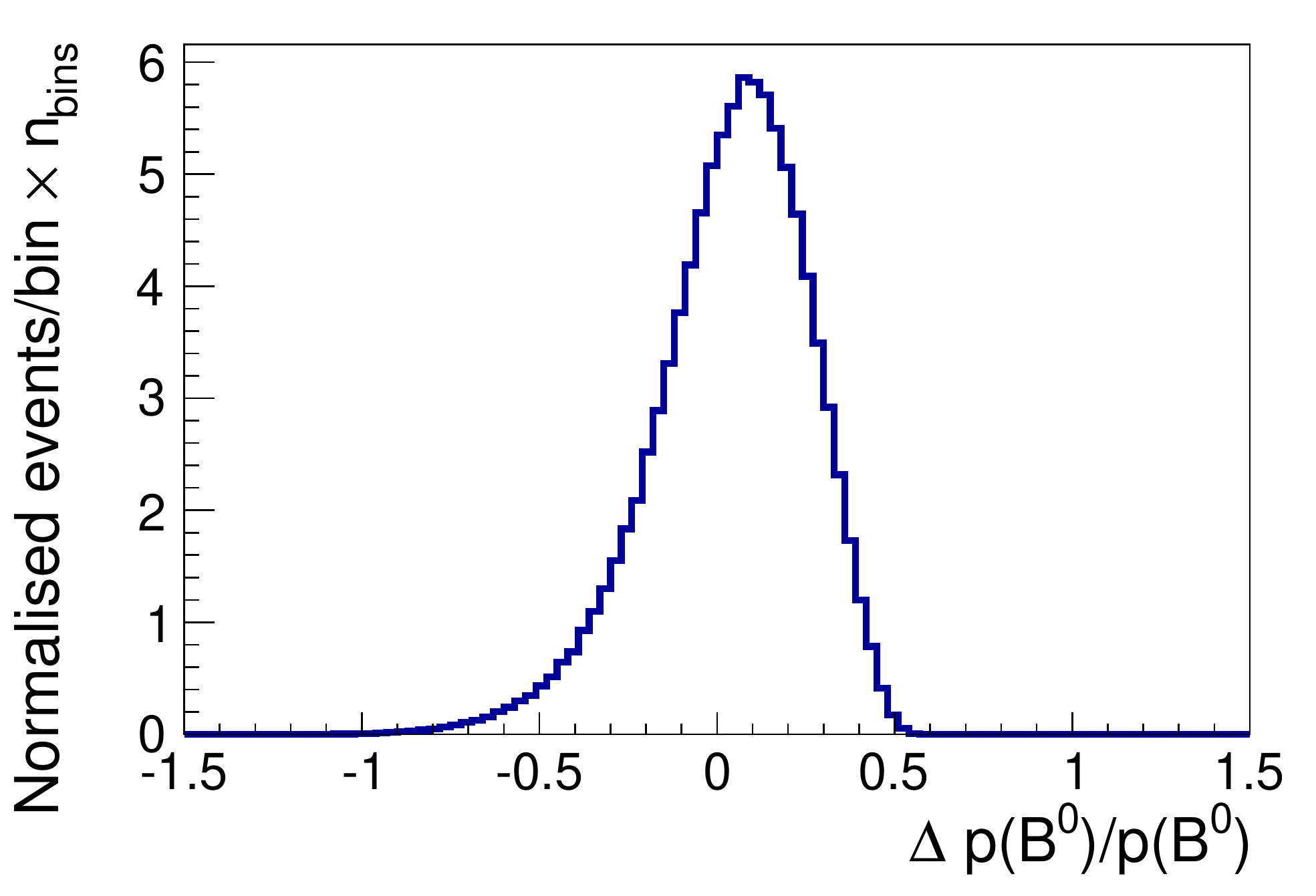}
\caption{Relative resolution on the \Bbar momentum magnitude for \BtoDstartaunutomu decays, obtained using equal velocity algorithm.\label{fig:B2Dsttaunumu_B_P_res}}
\end{figure}

\begin{figure}
\centering
\includegraphics[scale=0.7]{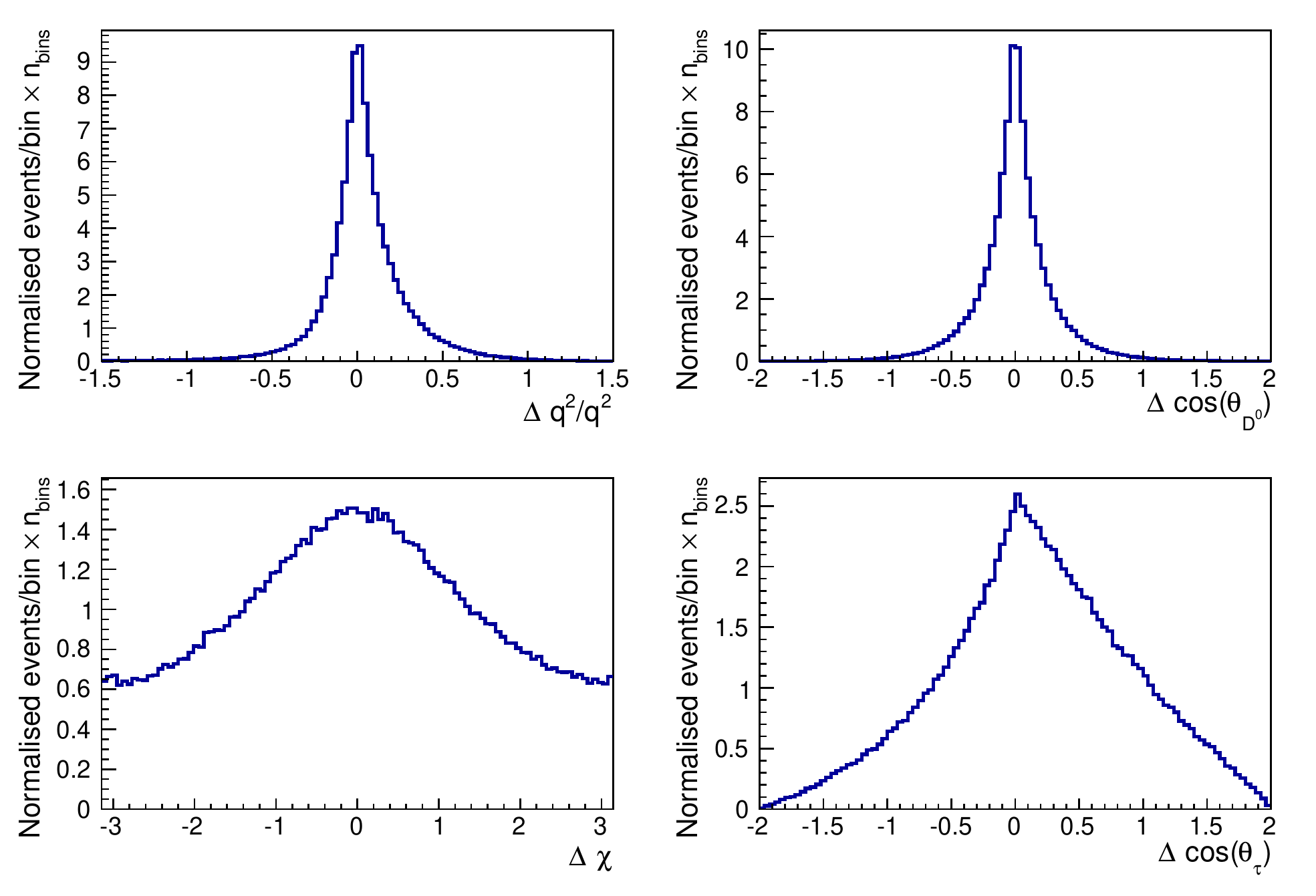}
\caption{Resolution on the phase space variables for \BtoDstartaunutomu decays, obtained with equal velocity algorithm. \label{fig:B2Dsttaunumu_distributions_res}}
\end{figure}

For \BtoDstartaunutomu decays, no information on the $\tau^-$ decay vertex is available and the equal velocity algorithm is applied. The relative resolution on the \Bbar momentum magnitude is shown in figure~\ref{fig:B2Dsttaunumu_B_P_res} and phase space variables resolutions are reported in figure~\ref{fig:B2Dsttaunumu_distributions_res}. The muon momentum is taken as tau lepton momentum for computing $\chi$ and \costhetal variables. Comparing to the tau lepton hadronic decay channel, the distributions are similar to the more precise resolutions of the full reconstruction algorithm rather than to those obtained with the equal velocity algorithm. Thus, the knowledge of the tau lepton flight direction in the three pion decay mode is not able to add significant information to the decay reconstruction due to the increased ambiguity in the \Bbar momentum determination.


Summarizing, \costhetaD and $\chi$ resolution distributions have been shown to be symmetric and unbiased for all the \BtoDstarlnu decay channels, and the related physical quantities are therefore suitable to be measured even at hadron collider experiments, making use of the presented reconstruction algorithms only. On the contrary, \costhetal resolution distributions have been found to be biased for $\tau$ lepton decay channels. The measurement of observables depending on \costhetal would therefore require special care and it is not further considered in this article.

\subsection{Extraction of angular distributions from the template fit selection}
\label{sec:extraction}
The selection of \BtoDstarlnu decays is a challenging task, especially at hadronic colliders. The impossibility of reconstructing all the final-state particles prevents the direct use of invariant masses as discriminating variables and makes different decays with similar topology but additional unreconstructed particles difficult to distinguish from \BtoDstarlnu transitions. In fact, besides discriminating muon from tau lepton decay modes, \BtoDstarlnu decays must be separated from \Bbar decays to \Dz, \Dp and other higher mass charm meson resonances $D^{**}$ and \Bbar decays to double charm resonances in which one has a semileptonic decay. This is usually achieved by means of a template fit to a set of discriminating variables, in which shapes for each decay type are mainly determined from simulation~\cite{Aaij:2015yra,Aaij:2017deq}.

The extraction of \BtoDstarlnu distributions from the fit results can be done straightforwardly by means of the \splot statistical tool~\cite{Pivk:2004ty} only for angular variables independent from the discriminating ones. In this way the distributions are derived using no a priori information about them, but only from the discriminating variables. Distributions which are correlated with the discriminating variables can also be obtained in principle, but since they will depend directly on the construction of the template distributions, their extraction would need a specific statistical treatment and they would be more sensitive to fit-related systematic uncertainties.

\begin{figure}
\centering
\includegraphics[scale=0.7]{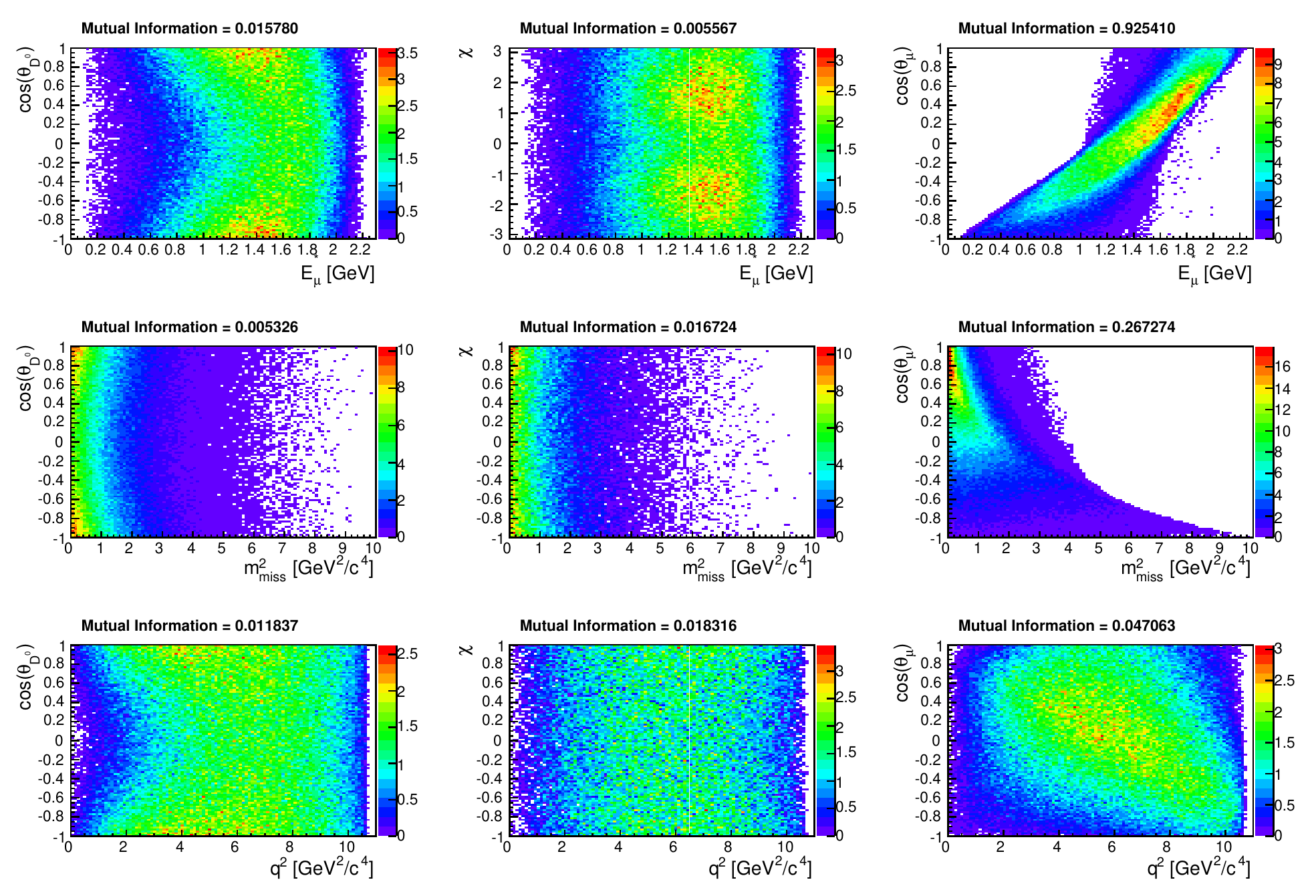}
\caption{Mutual information between angular and discriminating variables, for \BtoDstarmunu decays. \label{fig:B2Dstmunu_correlations}}
\end{figure}

\begin{figure}
\centering
\includegraphics[scale=0.7]{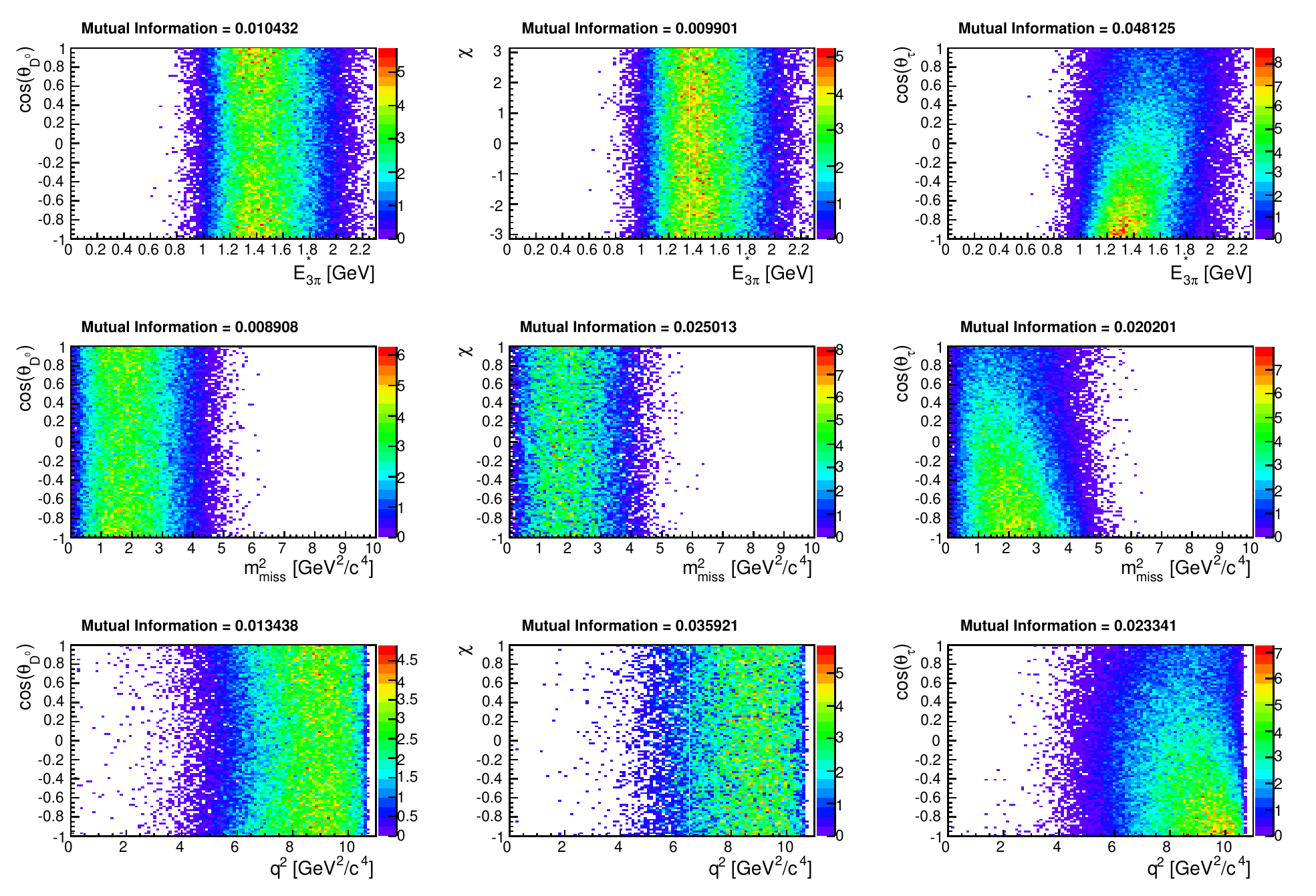}
\caption{Mutual information between angular and discriminating variables, for \BtoDstartaunutopi decays. \label{fig:B2Dsttaunupi_correlations}}
\end{figure}

\begin{figure}
\centering
\includegraphics[scale=0.7]{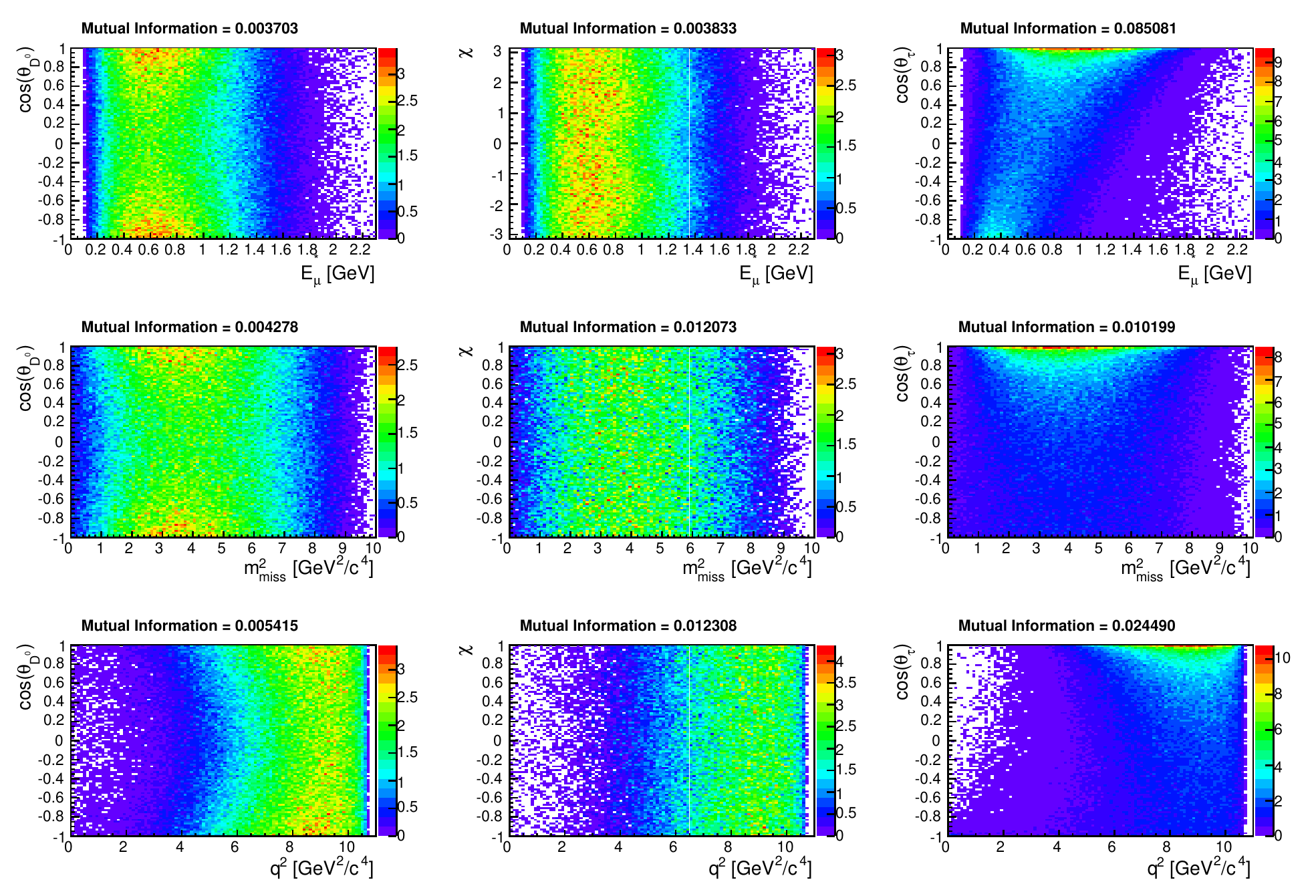}
\caption{Mutual information between angular and discriminating variables, for \BtoDstartaunutomu decays. \label{fig:B2Dsttaunumu_correlations}}
\end{figure}

The possibility of deriving \BtoDstarlnu angular distributions from a realistic selection is checked by evaluating their correlations, computed as mutual information,\footnote{The mutual information between two random variables $X$, $Y$, given their joint and marginalized probability distributions $p(X,Y)$ and $p(X)$, $p(Y)$, defined as
\begin{equation}
I(X:Y) = \sum_{X,Y} p(X,Y) \log\frac{p(X,Y)}{p(X)p(Y)},
\end{equation}
is sensitive to any form of relationship.} with the set of the three discriminating variables used in \cite{Aaij:2015yra}, in which the detectable part of the leptonic decay, $\lambda = \mu^-$ or $\lambda = \pi^+\pi^-\pi^-$, is used: the missing mass of the decay
\begin{equation}
m^2_{miss}=(p(\Bbar)-p(\Dstar)-p(\lambda))^2,
\end{equation}
the energy of the $\lambda$ system in the \Bbar rest frame $E^*_{\lambda}$, and $q^2$, where the \Bbar rest frame is estimated using the equal velocity algorithm. Correlation plots are presented in figure~\ref{fig:B2Dstmunu_correlations}, figure~\ref{fig:B2Dsttaunupi_correlations} and figure~\ref{fig:B2Dsttaunumu_correlations} for \BtoDstarmunu, \BtoDstartaunutopi and \BtoDstartaunutomu events, respectively. Since the discriminating variables depend on the leptonic part of the decay, correlations for \costhetaD and $\chi$ variables are found to be negligible; for \costhetal correlations are high for the muon decay mode and small for the tau lepton one, because in the latter case the relationship is blurred by the extra neutrinos coming from the $\tau^-$ decay.

Detector reconstruction and event selection may introduce additional correlations between discriminating and angular variables, but efficiency corrections are able to subtract these effects. Per-event efficiency corrections are routinely applied in many particle physics analyses, usually obtained from high-statistics simulation samples.

Thanks to their small correlations with the discriminating variables, \costhetaD and $\chi$ distributions can be extracted directly from the template fit using the \splot statistical technique, allowing related observable measurements to be performed on ``signal-only'' \costhetaD and $\chi$ distributions.


\section{The \BtoDstarlnu decay distribution}
\label{sec:decay_distribution}
Maximum information about the \BtoDstarlnutod decay is obtained from the fully differential decay distribution~\cite{Duraisamy:2013kcw}
\begin{align}
\frac{d^4\Gamma}{dq^2 d\costhetaD d\costhetal d\chi} &= \frac{9}{32\pi} \text{NF} \bigg[ \cosqthetaD \left( V_1^0 + V_2^0 \cos2\thetal + V_3^0 \costhetal \right) \nonumber\\
&+ \sin^2\thetaD \left( V_1^T + V_2^T \cos2\thetal + V_3^T \costhetal \right) \nonumber\\
&+ V_4^T \sin^2\thetaD \sin^2\thetal\cos 2\chi + V_1^{0T}\sin 2\thetaD\sin 2\thetal \cos\chi \nonumber\\
&+ V_2^{0T}\sin 2\thetaD\sin \thetal \cos\chi + V_5^T \sin^2\thetaD \sin^2\thetal\sin 2\chi \nonumber\\
&+ V_3^{0T}\sin 2\thetaD\sin \thetal \sin\chi + V_4^{0T}\sin 2\thetaD\sin 2\thetal \sin\chi \bigg],
\label{eq:BtoDstarlnu_full_distribution}
\end{align}
in which the dependence on the angular variables \costhetaD, \costhetal and $\chi$ has been made explicit. The decay is described by twelve angular coefficient functions $V_i$, dependent on couplings, hadronic form factors and $q^2$; NF is a $q^2$-dependent normalization term. The angular coefficients are labelled according to the \Dstar helicity combinations on which they depend: longitudinal ($V_i^0$), transverse ($V_i^T$) or mixed ($V_i^{0T}$).

The \CP-conjugate \BtoDstarlnuCP decay distribution follows from the application of the \CP transformation to equation~\eqref{eq:BtoDstarlnu_full_distribution}: the angles are now defined with respect to $l^+$ and \Dstarbar antiparticles, and the inversion of the momenta correspond to a transformation $\chi \to -\chi$ and $\thetal \to \theta_l+\pi$,
\begin{align}
\frac{d^4\bar{\Gamma}}{dq^2 d\costhetaD d\costhetal d\chi} &= \frac{9}{32\pi} \text{NF} \bigg[ \cosqthetaD \left( \bar{V}_1^0 + \bar{V}_2^0 \cos2\thetal - \bar{V}_3^0 \costhetal \right) \nonumber\\
&+ \sin^2\thetaD \left( \bar{V}_1^T + \bar{V}_2^T \cos2\thetal - \bar{V}_3^T \costhetal \right) \nonumber\\
&+ \bar{V}_4^T \sin^2\thetaD \sin^2\thetal\cos 2\chi + \bar{V}_1^{0T}\sin 2\thetaD\sin 2\thetal \cos\chi \nonumber\\
&- \bar{V}_2^{0T}\sin 2\thetaD\sin \thetal \cos\chi - \bar{V}_5^T \sin^2\thetaD \sin^2\thetal\sin 2\chi \nonumber\\
&+ \bar{V}_3^{0T}\sin 2\thetaD\sin \thetal \sin\chi - \bar{V}_4^{0T}\sin 2\thetaD\sin 2\thetal \sin\chi \bigg].
\label{eq:BtoDstarlnu_full_distribution_CP}
\end{align}
Angular terms proportional to $\sin\chi$ and $\sin2\chi$ are sensitive to \CP-violation, being produced in the interference between amplitudes having different \CP-violating weak phases. The associated coefficients, $V_5^T$, $V_3^{0T}$ and $V_4^{0T}$, are practically zero in the Standard Model~\cite{Duraisamy:2013kcw}; therefore a non-zero measurement of these quantities would be a clear sign of beyond the Standard Model physics.

Due to the experimentally available limited statistics, it is useful to integrate the fully differential decay distribution described by equation~\eqref{eq:BtoDstarlnu_full_distribution} to obtain observables retaining specific parts of the decay information. An overview of interesting observables defined for the \BtoDstarlnu decay distribution can be found in~\cite{Duraisamy:2013kcw,Colangelo:2018cnj,Becirevic:2016hea}; the following section will focus on observables constructed from \costhetaD and $\chi$ variables, the most suitable quantities to be measured according to the simulation study presented in section~\ref{sec:reconstruction}.

\subsection{Integrated distributions and observables}
\label{sec:integrated_distributions}
According to the study detailed in section~\ref{sec:resolutions}, the best resolution is attained on the polar angle of the \Dz meson in the \Dstar helicity frame, \costhetaD. The singly-differential distribution over \costhetaD, obtained integrating the complete decay distribution described by equation~\eqref{eq:BtoDstarlnu_full_distribution} over all but the \costhetaD variable, is
\begin{equation}
\frac{d\Gamma}{d\costhetaD} = \frac{3}{4} \left( 2\fl \cosqthetaD + \ft \sin^2\thetaD \right),
\end{equation}
in which \fl and \ft represent the $q^2$-integrated longitudinal and transverse polarization fractions of the \Dstar meson, satisfying $\fl+\ft=1$; the distribution takes the form of a second-order polynomial in \costhetaD depending on one single observable \fl,
\begin{equation}
\frac{d\Gamma}{d\costhetaD} = \frac{3}{4} \left[1-\fl + \left(3\fl-1 \right) \cosqthetaD \right].
\label{eq:FL_distribution}
\end{equation}
The \Dstar longitudinal polarization fraction is sensitive to scalar and tensor New Physics contributions to the $b\to c$ quark transition effective Hamiltonian, rather than to vector or axial-vector terms~\cite{Duraisamy:2013kcw,Alok:2016qyh}. Its ability to constrain New Physics contribution has been recently considered in~\cite{Blanke:2018yud,Iguro:2018vqb,Bhattacharya:2018kig}.

Observables derived from $\chi$-dependent decay distributions are especially interesting being clean probes for New Physics \CP-violation. Trigonometric functions of the $\chi$ angle can be expressed in terms of the unit vectors orthogonal to the \Dstar and $l^{-}\nul$ decay planes in the \Bbar meson rest frame,
\begin{equation}
\hat{n}_{D} = \frac{\hat{p}_{\Dz}\times \hat{p}_{\pi}}{|\hat{p}_{\Dz}\times \hat{p}_{\pi}|},\hspace{20pt}
\hat{n}_{W} = \frac{\hat{p}_{l^{-}}\times \hat{p}_{\nul}}{|\hat{p}_{l^{-}}\times \hat{p}_{\nul}|},\hspace{20pt}
\hat{n}_z = \left\lbrace 0,0,1 \right\rbrace = \frac{\hat{p}_{\Dz} + \hat{p}_{\pi}}{|\hat{p}_{\Dz} + \hat{p}_{\pi}|},
\end{equation}
as
\begin{equation}
\cos\chi = \hat{n}_{D} \cdot \hat{n}_{W}, \hspace{20pt} \sin\chi = \left( \hat{n}_{D} \times \hat{n}_{W} \right) \cdot \hat{n}_z,
\end{equation}
so that observables which are coefficients of $\sin\chi$ or $\sin2\chi$ can be extracted as triple-product asymmetries. This feature allows \CP-violating observables to be extracted by counting rather than by angular fits and will be exploited further on.

The singly-differential distribution over $\chi$ is obtained by integrating equation~\eqref{eq:BtoDstarlnu_full_distribution}
\begin{equation}
\frac{d\Gamma}{d\chi} = \frac{1}{2\pi} \left( 1+ \Ac \cos2\chi + \At \sin2\chi \right).
\label{eq:Ac_At_distribution}
\end{equation}
The \CP-violating \At observable is sensitive to vector and axial vector New Physics contributions but not to pseudoscalar ones~\cite{Duraisamy:2013kcw}. It depends linearly on $V^T_5$, while for the \CP-conjugated decay \Atbar depends on $-\bar{V}^T_5$, changing sign under \CP-transformation. The corresponding \CP-violating observable can be thus defined as
\begin{equation}
\Acp = \frac{\At + \Atbar}{2}.
\end{equation}
Exploiting the odd parity of the $\sin2\chi$ term, the \At observable can be isolated from the distribution described by equation~\eqref{eq:Ac_At_distribution} by defining the triple-product asymmetry
\begin{equation}
\TPA = \int \mathrm{sign}(\sin2\chi) \frac{d\Gamma}{d\chi} d\chi = \frac{2}{\pi} \At.
\label{eq:TPA1}
\end{equation}
The sum of \TPA asymmetries measured for the two \CP-conjugated decays still represent a \CP-violating observable.

Terms proportional to $\sin\chi$ in the full decay distribution are multiplied by $\sin2\thetaD$ and integrate to zero under $\int d\costhetaD$. The triple-product asymmetry defined as
\begin{equation}
\TPAz = \int \mathrm{sign}(\sin\chi) \frac{d\Gamma}{d\chi} d\chi = 0,
\label{eq:TPA0}
\end{equation}
is zero even in presence of New Physics, being this angular dependence related to the spin structure of the \BtoDstarlnu decay, in which the \Dstar meson has spin one. The measurement of \TPAz is therefore a useful cross-check for the triple-product asymmetry measurement, allowing to assess possible biases or contamination from $\bar{\B}\to\Dz\pi l^-\bar{\nu}_{l}$ events in which the $\Dz\pi$ comes from a spin zero resonance decay, like the $D^{*+}_0(2400)$, or from a non-resonant system~\cite{Becirevic:2016hea}.

Observables related to the $\sin\chi$ terms of the decay distribution can be extracted from the $\chi$-dependent angular distribution defined as
\begin{equation}
\frac{d\Gamma^{(2)}}{d\chi} = \int \mathrm{sign}(\costhetaD) \frac{d\Gamma}{d\costhetaD d\chi} d\costhetaD = \frac{1}{4} \left( \Acc \cos\chi + \Att \sin\chi \right).
\label{eq:chi_gamma2_distribution}
\end{equation}
The \CP-violating \Att observable is sensitive to all vector, axial-vector and pseudoscalar couplings~\cite{Duraisamy:2013kcw}. It depends linearly on $V^{0T}_3$, while for the \CP-conjugated decay \Attbar depends on $\bar{V}^{0T}_3$, not changing sign under \CP-transformation. The corresponding \CP-violating observable is therefore
\begin{equation}
\Acpp = \frac{\Att - \Attbar}{2}.
\end{equation}
Starting from the distribution reported in equation~\eqref{eq:chi_gamma2_distribution}, a triple-product asymmetry equivalent to the \Att observable can be defined as
\begin{equation}
\TPAA = \int \mathrm{sign}(\sin\chi) \frac{d\Gamma^{(2)}}{d\chi} d\chi = \Att.
\label{eq:TPA2}
\end{equation}
The difference between \TPAA asymmetries measured for the two \CP-conjugated decays represents a \CP-violation observable.

\section{Measurement method for \BtoDstarlnu decay distribution observables}
\label{sec:measurement}
The non-negligible width of the resolution on the angular variables, studied in section~\ref{sec:resolutions}, must be taken into account when measuring the corresponding observables, which can be biased from their actual value. In section~\ref{sec:fl_measurement}, it is shown how the \Dstar longitudinal polarization can be extracted from maximum-likelihood fits to simulated \BtoDstarlnu events by parametrizing the detector response in \costhetaD and as a function of \fl via a polynomial expansion. This way, the non-negligible experimental resolution effect is subtracted, and the measured values are found compatible with the generated ones. The loss of sensitivity due to the experimental resolution is evaluated. Maximum-likelihood fits have been performed using the ROOFIT package~\cite{Verkerke:2003ir}.

The same method is then applied for the extraction of \Ac and \At observables, section~\ref{sec:at_measurement}, but found to be successful only for \BtoDstarmunu decays, due to the too large uncertainties associated to the $\chi$ angle reconstruction for tau lepton decay modes.

Section~\ref{sec:TPA_measurement} deals with triple-product asymmetries, which can be measured just by counting. In this case, the simulation is used to determine the proportionality factor between the \CP-violating observables and the associated reconstructed triple-product asymmetry, allowing to correct for the experimental resolution and to quantify the associated loss in precision.

\subsection{\Dstar longitudinal polarization}
\label{sec:fl_measurement}
As a first step, a per-event weight is assigned to simulated \BtoDstarlnu decays in order to obtain a flat distribution in the generated \costhetaD values, for correcting the distortion due the applied geometry and selection requirements. Different longitudinal polarizations are generated by applying another per-event, polarization dependent, weight such that the generated \costhetaD distribution reproduces equation~\eqref{eq:FL_distribution} for each \fl value. Both weights are normalized in such a way that for each \fl value the mean of the weights is one.

The two per-event weights are multiplied together, assuming the detector efficiency correction is independent of \fl. This assumption has been checked to be valid for the presented simulation study. In a real-case analysis, the generation of \BtoDstarlnu events with varying longitudinal polarization should be done before applying the detector reconstruction, so that detector efficiency effects can be taken into account as a function of \fl.

Simulated events are then divided in two samples: a test sample reproducing \BtoDstarlnu reconstructed decays with different \Dstar longitudinal polarizations, and a second used to derive a Legendre polynomial expansion in \costhetaD and \fl. This expansion is used as fit model to extract \fl from a maximum-likelihood fit of the test sample. The orthogonality and completeness of Legendre polynomials $L(x,i)$ is exploited to expand the reconstructed decay distribution in \costhetaD and \fl as
\begin{equation}
p(\costhetaD,\fl) = \sum_{i,j} c_{i,j} L(\costhetaD,i)L(\fl,j),
\label{eq:poly_exp_fl}
\end{equation}
in which the coefficients $c_{i,j}$ are determined as
\begin{equation}
c_{i,j} = \sum_{n=0}^{N} w_n(\costhetaD,\fl) \left(\frac{2i+1}{2}\right)\left(\frac{2j+1}{2}\right) L(\costhetaD,i)L(\fl,j),
\end{equation}
and $w_n(\costhetaD,\fl)$ is the product of the two per-event weights applied. Given the simple dependencies, quadratic in \costhetaD and linear in \fl, only Legendre polynomials up to the second order are sufficient to approximate the decay distribution. The use of a simple parametrization makes the maximum-likelihood fit of the decay distribution fast and robust.

The test samples contain, by choice, ten thousand \BtoDstarlnu events per decay mode, while the other samples are five times larger than the test one. This is equivalent to assume that, in a real measurement, the statistics of the simulation sample employed to derive the polynomial expansion is larger enough with respect to the data sample.

The sensitivity to the \Dstar longitudinal polarization is studied by fitting the test samples using directly the angular distribution equation~\eqref{eq:FL_distribution} or the polynomial expansions equation~\eqref{eq:poly_exp_fl} for the three considered \BtoDstarlnu decays. The measured polarizations are reported in table~\ref{tab:FL_measurement}. Ideal \fl measurements are simulated by fitting the angular distribution described by equation~\eqref{eq:FL_distribution} to a toy sample generated from the same distribution for varying \fl values, with the same number of events of the test samples. These correspond to measurements made by a detector with perfect \costhetaD resolution, taken as reference to evaluate the decrease in precision due to the reconstruction algorithms employed. Results of these ideal measurements are reported in the last row of table~\ref{tab:FL_measurement}.

Longitudinal polarizations extracted using the true angular distributions are clearly biased towards values for which the \costhetaD distribution is flatter (it is uniform for $\fl=1/3$). Polynomial expansions allow to correctly measure the generated values within the uncertainties resulting from the maximum-likelihood fit. The precision for different \fl values with respect to the ideal case decreases by a factor 1.4--1.9 for the muon mode and a factor 1.5--2 for the \BtoDstartaunutomu decay. The precision is therefore similar for muon and tau lepton decay modes, as expected since the \costhetaD variable does not directly depend on the leptonic part of the decay. The exploitation of the tau lepton decay vertex information in the \BtoDstartaunutopi decay reconstruction does not increase the precision on \fl, rather, a larger uncertainty is observed for this mode.

\begin{table}
\centering
\begin{tabular}{lccc}
\hline
\fl (gen) & 10 & 50 & 90\\
\hline
\fl (\BtoDstarmunu, true) & 12.65 $\pm$  0.60 & 41.61 $\pm$  0.76 & 71.36 $\pm$  0.71\\
\fl (\BtoDstartaunu ($3\pi$), true) & 16.79 $\pm$  0.65 & 41.37 $\pm$  0.76 & 66.29 $\pm$  0.73 \\
\fl (\BtoDstartaunu ($\mu$), true) & 16.58 $\pm$  0.65 & 44.05 $\pm$  0.77 & 71.52 $\pm$  0.70 \\
\fl (\BtoDstarmunu, expansion) & 10.18 $\pm$  0.84 & 50.42 $\pm$  1.06 & 91.76 $\pm$  0.99 \\
\fl (\BtoDstartaunu ($3\pi$), expansion) & 10.49 $\pm$  1.06 & 50.58 $\pm$  1.23 & 90.81 $\pm$  1.18\\
\fl (\BtoDstartaunu ($\mu$), expansion) & 9.82 $\pm$  0.96 & 50.29 $\pm$  1.13 & 90.72 $\pm$  1.04 \\
\hline
\fl (gen, true) & 10.13 $\pm$  0.58 & 50.24 $\pm$  0.76 & 90.10 $\pm$  0.52\\
\hline
\end{tabular}
\caption{Measured \Dstar longitudinal polarization (in \%) by fitting the true angular distribution equation~\eqref{eq:FL_distribution} or the polynomial expansions equation~\eqref{eq:poly_exp_fl} to the \BtoDstarlnu test samples for varying generated \fl values; the last row reports the ideal measurements obtained by fitting the true angular distribution to a toy sample generated from the same distribution with the same number of events of the test sample. \label{tab:FL_measurement}}
\end{table}

According to this simulation study, the \Dstar polarization fraction of \BtoDstarlnu decays is measurable with the sole use of the employed reconstruction algorithm, with a maximum penalty in sensitivity of a factor 2. This permits an additional search for New Physics in \BtoDstarlnu decays complementary to the already measured \RDstar ratio.

\subsection{An attempt to directly measure the \At and \Ac observables}
\label{sec:at_measurement}
A simulation study analogous to the one set for the \fl measurement is performed to check the possibility to simultaneously measure the \At and \Ac observables related to the distribution reported in equation~\eqref{eq:Ac_At_distribution}. This case is more difficult partly because of the larger resolution on the $\chi$ angle, especially for the tau lepton decay mode, partly because this angular distribution is characterized by fast oscillations ($\cos2\chi$ and $\sin 2\chi$ terms) more sensitive to reconstruction inaccuracies.

A first study is carried out assuming that the $\chi$ distribution has a simpler form,
\begin{equation}
\frac{d\Gamma}{d\chi} \equiv \frac{1}{2\pi} \left( 1+ \Acz \cos\chi + \Atz \sin\chi \right),
\label{eq:Acz_Atz_distribution}
\end{equation}
in which the $\chi$ oscillations are wider. As explained in section~\ref{sec:decay_distribution}, this angular dependence is absent from the actual \BtoDstarlnu distribution, so that the angular coefficients \Acz, \Atz do not correspond to \BtoDstarlnu angular observables. They are introduced with the purpose of testing the extraction method already applied to \fl. The fit model is derived from the reconstructed decay distribution by means of a polynomial expansion in $\chi$, \Acz and \Atz: Legendre polynomials are used for \Acz and \Atz, while a Fourier series\footnote{
A generic function defined over the range $\left[ -\pi,\pi \right]$ can be expanded as a linear combination of the orthonormal basis
\begin{equation*}
\bigg\lbrace\frac{1}{2\pi}, \frac{1}{\pi}\cos(nx), \frac{1}{\pi}\sin(nx)\bigg\rbrace.
\end{equation*}
}
up to $\cos2\chi$, $\sin 2\chi$ terms is employed for the $\chi$ angle,
\begin{equation}
p(\chi,\Acz,\Atz) = \sum_{i,j,k} c_{i,j,k} F(\chi,i)L(\Acz,j)L(\Atz,k),
\label{eq:poly_exp_atz}
\end{equation}
in which the coefficients $c_{i,j,k}$ are determined as
\begin{align}
c_{i,j,k} &= \sum_{n=0}^{N} w_n(\chi,\Acz,\Atz) \left(\frac{2j+1}{2}\right)\left(\frac{2k+1}{2}\right) \times F(\chi,i)L(\Acz,j)L(\Atz,k),
\end{align}
and $w_n(\chi,\Acz,\Atz)$ is the product of the two per-event weights applied.

The measured \Atz and \Acz values using the distribution equation~\eqref{eq:Acz_Atz_distribution} and the polynomial expansions equation~\eqref{eq:poly_exp_atz} are reported in tables~\ref{tab:Atz_measurement} and~\ref{tab:Acz_measurement}, respectively. Only results in which one of the two observables is zero are shown, since negligible differences in the observables extraction are seen when both \Atz and \Acz have non-zero values. Ideal measurements are also simulated as done for \fl. Only \BtoDstartaunutomu decays are considered for the tau lepton decay mode.

The polynomial expansions recover the generated values within uncertainties, with a precision on \Atz decreased by a factor 2--2.2 for the muon mode and 4.4--4.9 for the tau lepton mode, and a precision on \Acz decreased by a factor 1.8--2 for the muon mode and 5.5--6.3 for the tau lepton mode. As a result, the polynomial expansion method proves to be effective but the decrease in precision for the tau lepton decay mode is important to note.

The simulation study is repeated for \At and \Ac using the distribution described by equation~\eqref{eq:Ac_At_distribution} and an analogous
polynomial expansion. Unfortunately, the two observables are measurable only for the muon decay mode, the results of which are shown in tables~\ref{tab:At_measurement} and \ref{tab:Ac_measurement}, with precisions on \At and \Ac observables decreased by a factor 2.9--3.2 and 2.6--2.7, respectively. The measurement is not possible on the tau lepton decay mode because the large uncertainty in the reconstruction completely flattens the $\chi$ angle distribution.

\begin{table}
\centering
\begin{tabular}{lccc}
\hline
(\Acz,\Atz) (gen) & (0,0) & (0,50) & (0,-50)\\
\hline
\Atz (\BtoDstarmunu, true) & -2.22 $\pm$  1.41 & 20.90 $\pm$ 1.39 & -25.18 $\pm$ 1.37\\
\Atz (\BtoDstartaunu ($\mu$), true) & -0.60 $\pm$  1.41 & 11.42 $\pm$  1.41 & -12.39 $\pm$  1.39\\
\Atz (\BtoDstarmunu, exp.) & -3.66 $\pm$  2.94 & 49.56 $\pm$  2.89 & -51.60 $\pm$  2.86\\
\Atz (\BtoDstartaunu ($\mu$), exp.) & -3.58 $\pm$ 6.24 & 49.62 $\pm$  6.26 & -55.69 $\pm$  6.18\\
\hline
\Atz (gen, true) & -0.55 $\pm$ 1.42 & 49.55 $\pm$  1.28 & -50.45 $\pm$  1.27\\
\hline
\end{tabular}
\caption{Measured \Atz (in \%) fitting the angular distribution equation~\eqref{eq:Acz_Atz_distribution} or the polynomial expansions equation~\eqref{eq:poly_exp_atz} to the \BtoDstarlnu test samples for varying generated values; the last row reports the ideal sensitivity obtained from a toy sample generated from the true angular distribution with the same number of events of the test sample, fitted with the same distribution. \label{tab:Atz_measurement}}
\end{table}

\begin{table}
\centering
\begin{tabular}{lccc}
\hline
(\Acz,\Atz) (gen) & (0,0) & (50,0) & (-50,0)\\
\hline
\Acz (\BtoDstarmunu, true) & -0.85 $\pm$ 1.42 & 28.64 $\pm$  1.37 & -26.93 $\pm$  1.38\\
\Acz (\BtoDstartaunu ($\mu$), true) & -2.06 $\pm$  1.42 & 6.87 $\pm$  1.42 & -11.05 $\pm$ 1.42\\
\Acz (\BtoDstarmunu, exp.) & 2.55 $\pm$ 2.62 & 53.89 $\pm$  2.53 & -48.92 $\pm$  2.56\\
\Acz (\BtoDstartaunu ($\mu$), exp.) & -5.47 $\pm$ 7.84 & 44.45 $\pm$ 7.91 & -56.49 $\pm$ 8.05\\
\hline
\Acz (gen, true) & 0.62 $\pm$ 1.41 & 50.50 $\pm$  1.26 & -49.49 $\pm$ 1.27\\
\hline
\end{tabular}
\caption{Measured \Acz (in \%) fitting the angular distribution equation~\eqref{eq:Acz_Atz_distribution} or the polynomial expansions equation~\eqref{eq:poly_exp_atz} to the \BtoDstarlnu test samples for varying generated values; the last row reports the ideal sensitivity obtained from a toy sample generated from the true angular distribution with the same number of events of the test sample, fitted with the same distribution. \label{tab:Acz_measurement}}
\end{table}

\begin{table}
\centering
\begin{tabular}{lcccc}
\hline
(\Ac,\At) (gen) & (0,0) & (0,50) & (0,-50)\\
\hline
\At (\BtoDstarmunu, true) & -0.38 $\pm$  1.42 & 16.81 $\pm$  1.40 & -17.66 $\pm$  1.40\\
\At (\BtoDstarmunu, exp.) & -1.98 $\pm$  4.05 & 47.10 $\pm$  3.99 & -51.46 $\pm$  4.03\\
\hline
\At (gen, true) & 1.25 $\pm$ 1.41 & 51.00 $\pm$ 1.26 & -48.98 $\pm$  1.28 \\
\hline
\end{tabular}
\caption{Measured \At (in \%) fitting the angular distribution equation~\eqref{eq:Ac_At_distribution} or the polynomial expansions equation~\eqref{eq:poly_exp_atz} to the \BtoDstarmunu test sample for varying generated values; the last row reports the ideal sensitivity obtained from a toy sample generated from the true angular distribution with the same number of events of the test sample, fitted with the same distribution.\label{tab:At_measurement}}
\end{table}

\begin{table}
\centering
\begin{tabular}{lcccc}
\hline
(\Ac,\At) (gen) & (0,0) & (50,0) & (-50,0)\\
\hline
\Ac (\BtoDstarmunu, true) & -1.12 $\pm$  1.41 & 17.92 $\pm$ 1.39 & -20.59 $\pm$ 1.39\\
\Ac (\BtoDstarmunu, exp.) & 1.00 $\pm$  3.69 & 48.94 $\pm$ 3.38 & -53.14 $\pm$ 3.40\\
\hline
\Ac (gen, true) & 1.12 $\pm$ 1.41 & 50.79 $\pm$ 1.26 & -49.21 $\pm$ 1.28\\
\hline
\end{tabular}
\caption{Measured \Ac (in \%) fitting the angular distribution equation~\eqref{eq:Ac_At_distribution} or the polynomial expansions equation~\eqref{eq:poly_exp_atz} to the \BtoDstarmunu test sample for varying generated values; the last row reports the ideal sensitivity obtained from a toy sample generated from the true angular distribution with the same number of events of the test sample, fitted with the same distribution.\label{tab:Ac_measurement}}
\end{table}

The application of the polynomial expansion method is in principle effective for measuring $\chi$ angle related observables. In practice it is successful only for the \BtoDstarmunu decay mode, where \At and \Ac observables can be measured; for tau lepton decay modes the extraction is not possible due to both the larger resolution on the $\chi$ angle and the form of the expected decay distributions. The method has not been attempted for \Att and \Acc measurement because its application is complicated by the combined fit to \costhetaD and $\chi$ variables and the need for negative-valued fitting functions (following from the angular distribution described by equation~\ref{eq:chi_gamma2_distribution}), which prevent the use of the standard maximum-likelihood fitting technique.

An alternative method for the measurement of \CP-violating observables, relying on counting rather than fitting, is explored in the next section.

\subsection{Triple-product asymmetries}
\label{sec:TPA_measurement}
In section~\ref{sec:integrated_distributions} it was shown that \CP-violating observables related to $\chi$ angle decay distributions can be extracted by defining suitable triple-product asymmetries (TPAs). The imperfect reconstruction of the $\chi$ angle leads to an effective dilution of the asymmetries, but this experimental effect can still be subtracted exploiting \BtoDstarlnu simulated events, and in a simpler way than for decay angular distribution fits. Moreover, since the $\chi$ angle distribution is unbiased, a measured non-zero value for \CP-violating TPAs, even if not corrected for the experimental dilution, would anyway represent an observation of New Physics \CP-violation.

The subtraction of reconstruction effects consists in determining the relation between reconstructed TPAs and generated \CP-violating observables. The linear function $\mathrm{TPA} = f(\mathcal{A}_T)$ allows to infer $\mathcal{A}_T$ from the measured $\mathrm{TPA}$ with an uncertainty given by error propagation,
\begin{equation}
\sigma(\mathcal{A}_T) = \frac{\partial f^{-1}(\mathrm{TPA})}{\partial \mathrm{TPA}} \sigma(\mathrm{TPA}) = \left(\frac{\partial f(\mathcal{A}_T)}{\partial \mathcal{A}_T}\right)^{-1} \hspace{-12pt} \sigma(\mathrm{TPA})\equiv \kappa \sigma(\mathrm{TPA}),
\end{equation}
in which $\kappa$ represent the loss in sensitivity to $\mathcal{A}_T$ with respect to the uncertainty on the TPA.

The simulation study is set as follows. Simulated events are weighted to reproduce one of the $\chi$ angle decay distributions at generation-level, as a function of the \CP-violating observables. TPAs are built from the reconstructed value of the $\chi$ angle; for the distribution reported in equation~\eqref{eq:chi_gamma2_distribution} the $\sin2\thetaD$ dependence is included to take into account uncertainties in the \costhetaD sign determination. Three values for the corresponding \CP-conserving quantities $\mathcal{A}^{(i)}_C=0,\pm 1$ have been considered, but it is shown that they have no impact on the TPAs measurement. In fact, $\cos\chi$ and $\cos2\chi$ terms still integrate to zero when computing asymmetries using reconstructed angles, since the $\chi$ angle resolution distribution is not biased. The linear relation between reconstructed asymmetries and generated \CP-violating observables allows to correct for the dilution effects and to determine the decrease in precision from the inverse of the slope of the straight line.

\begin{figure}
\centering
\includegraphics[scale=0.24]{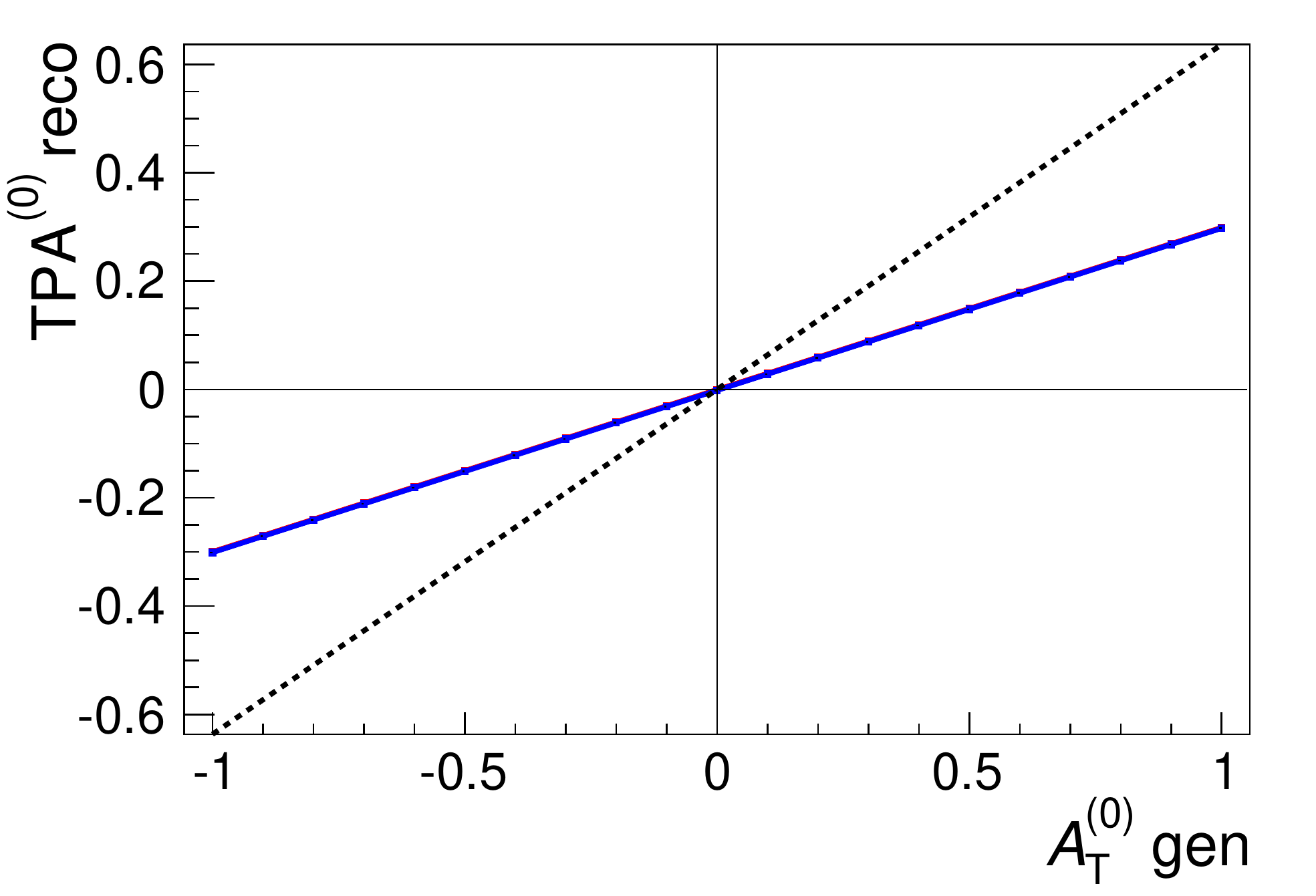}
\includegraphics[scale=0.24]{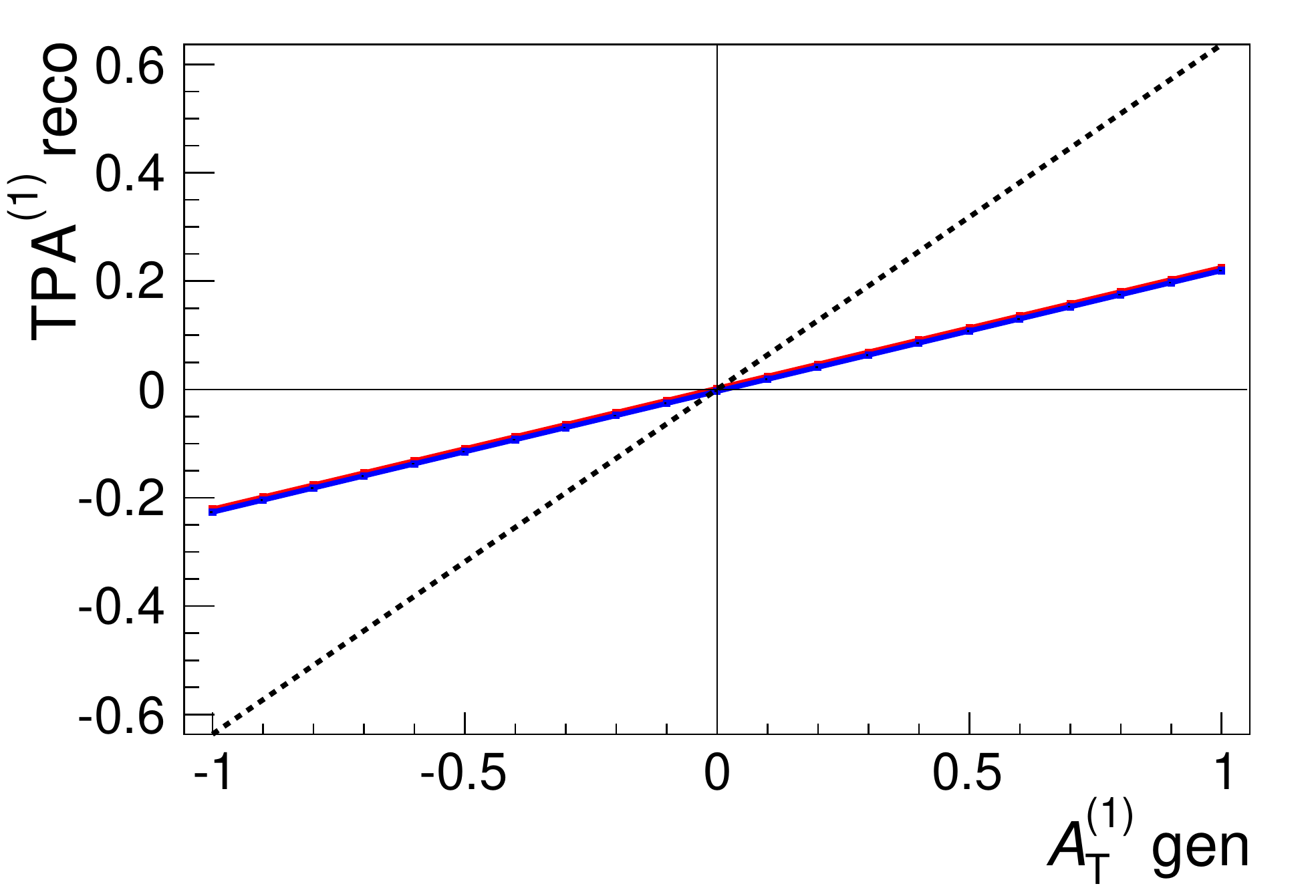}
\includegraphics[scale=0.24]{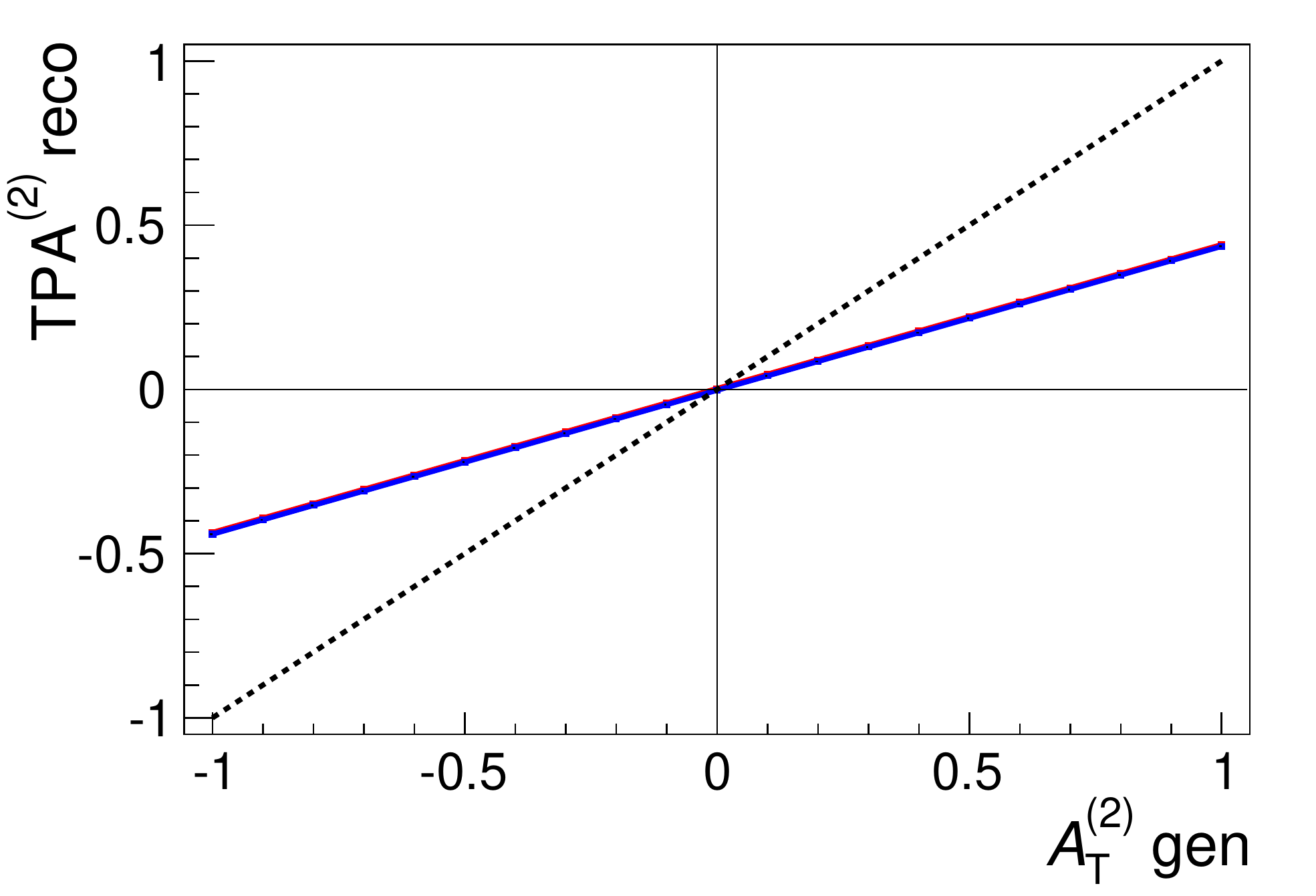}\\
\includegraphics[scale=0.24]{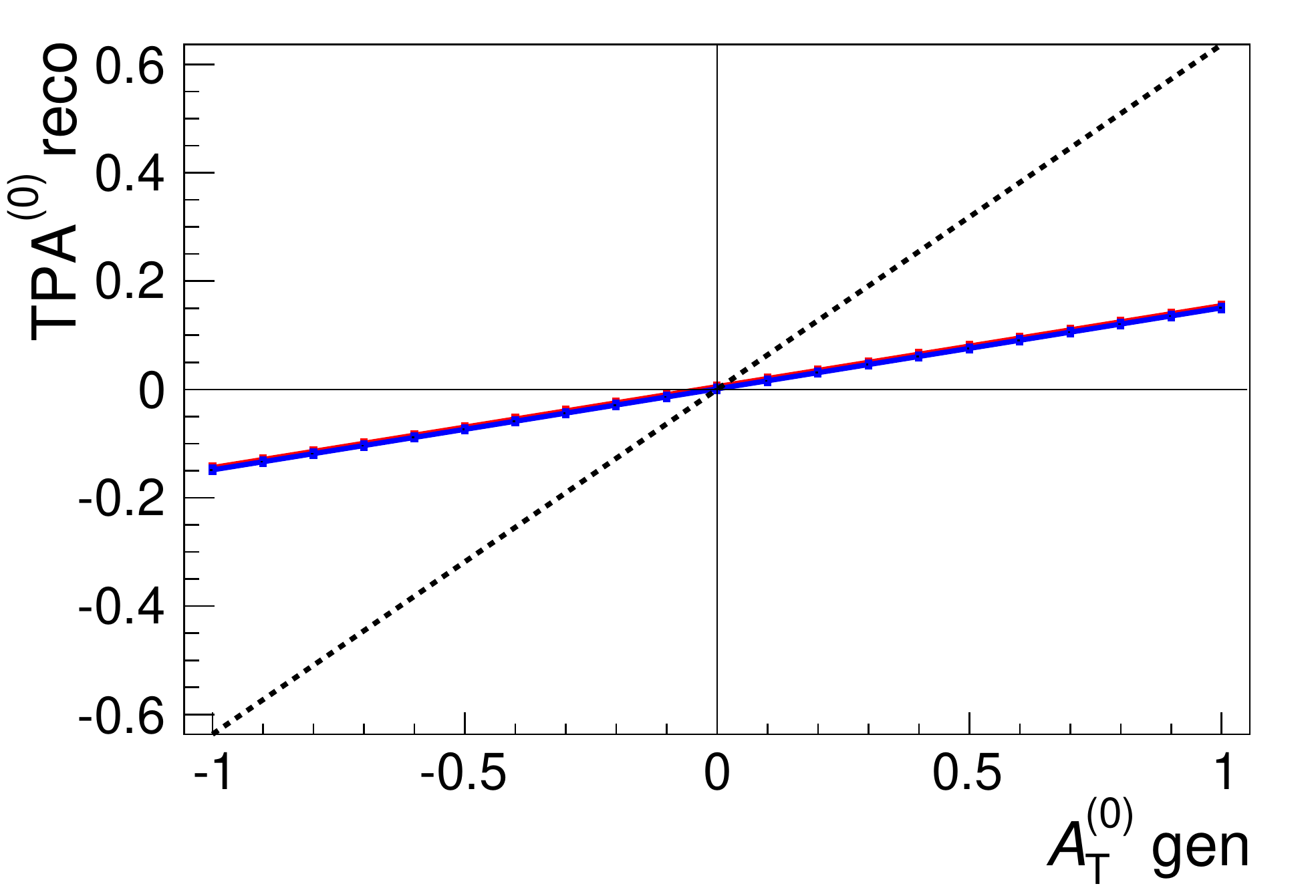}
\includegraphics[scale=0.24]{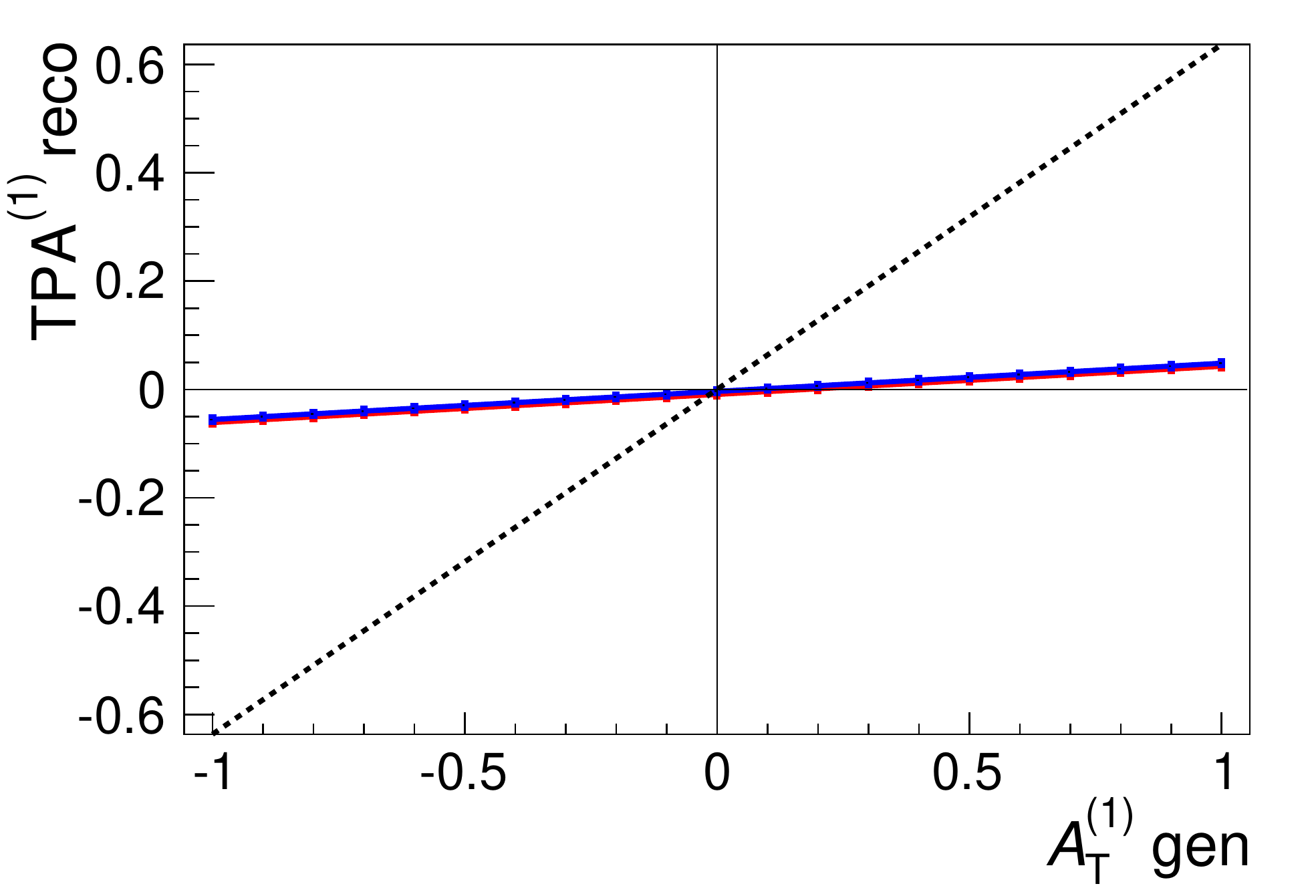}
\includegraphics[scale=0.24]{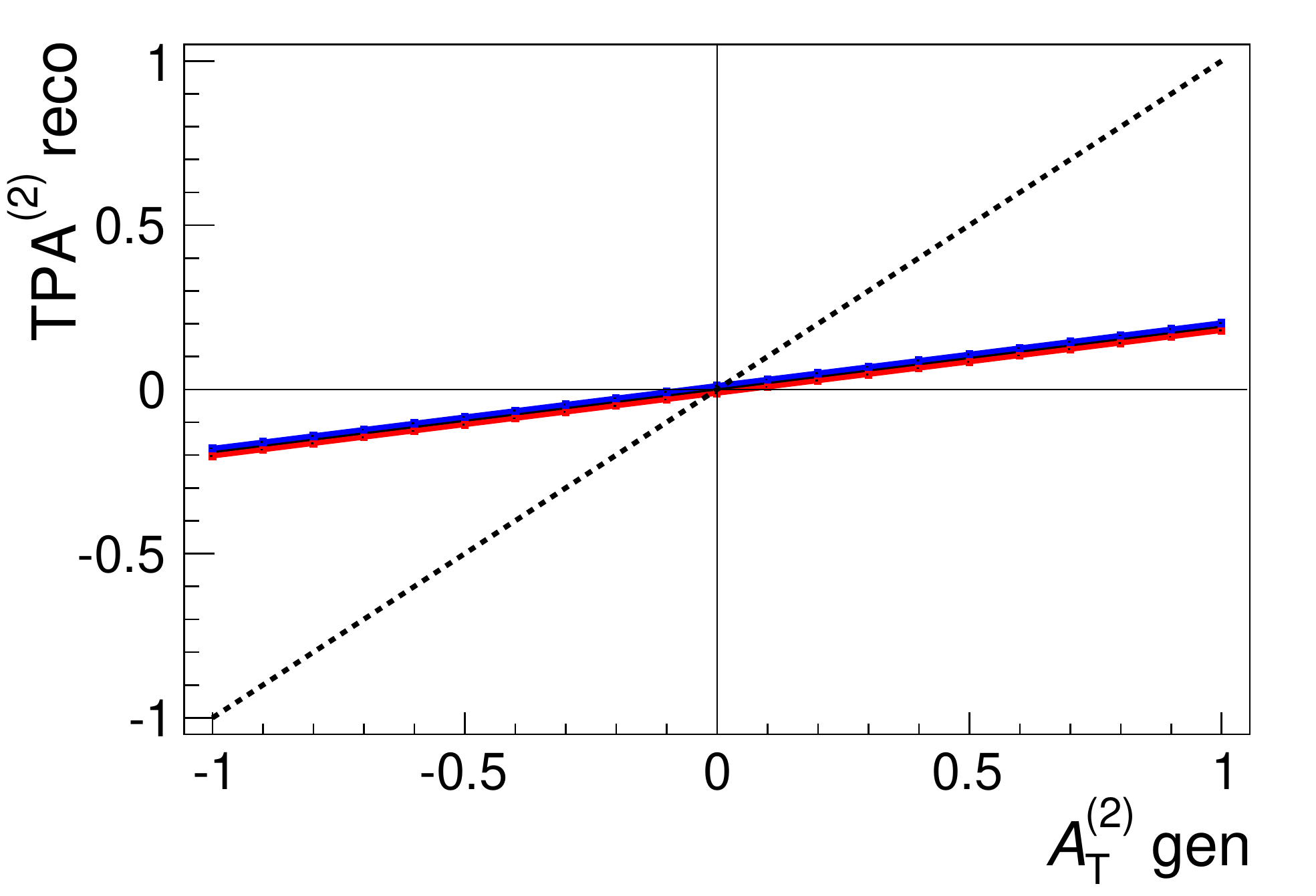}\\
\includegraphics[scale=0.24]{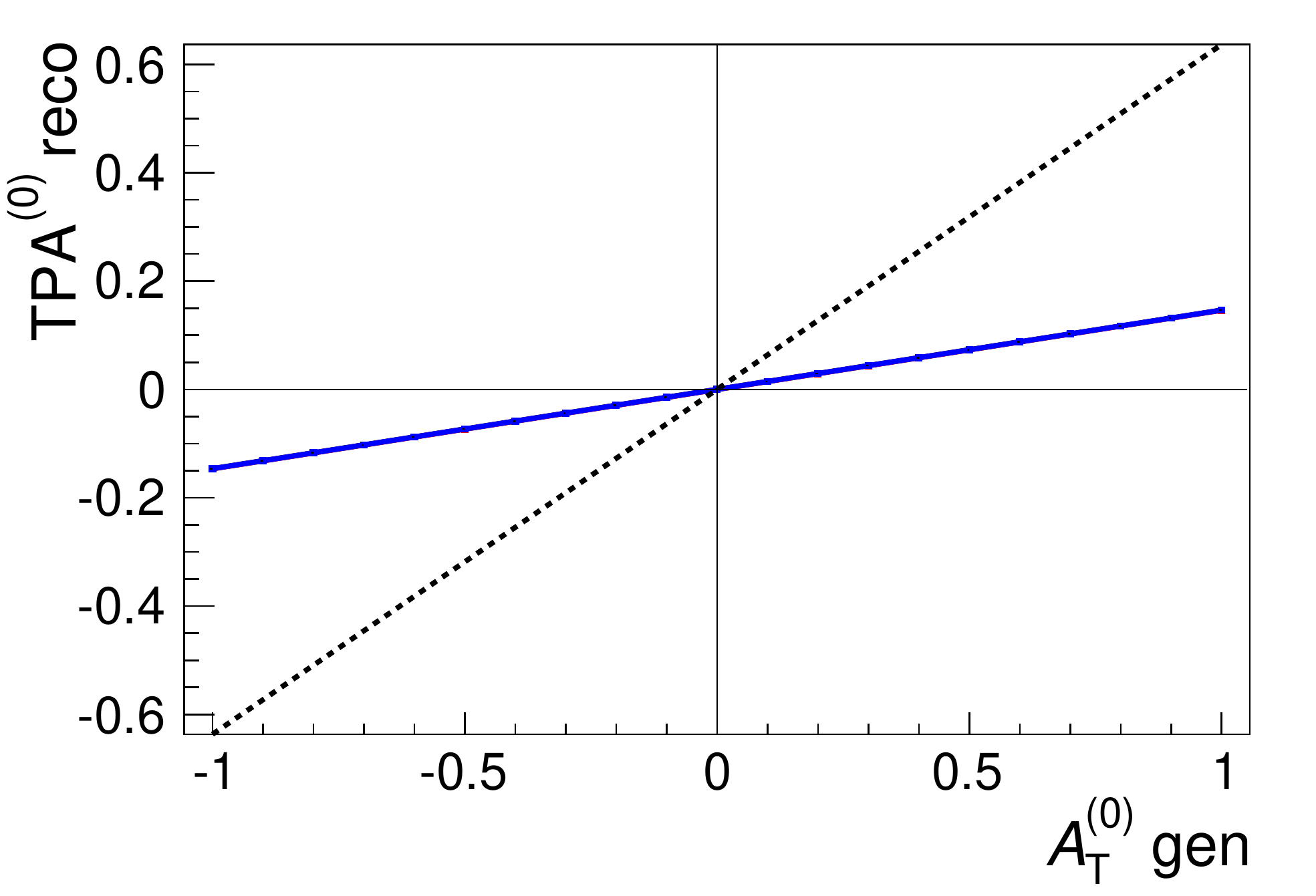}
\includegraphics[scale=0.24]{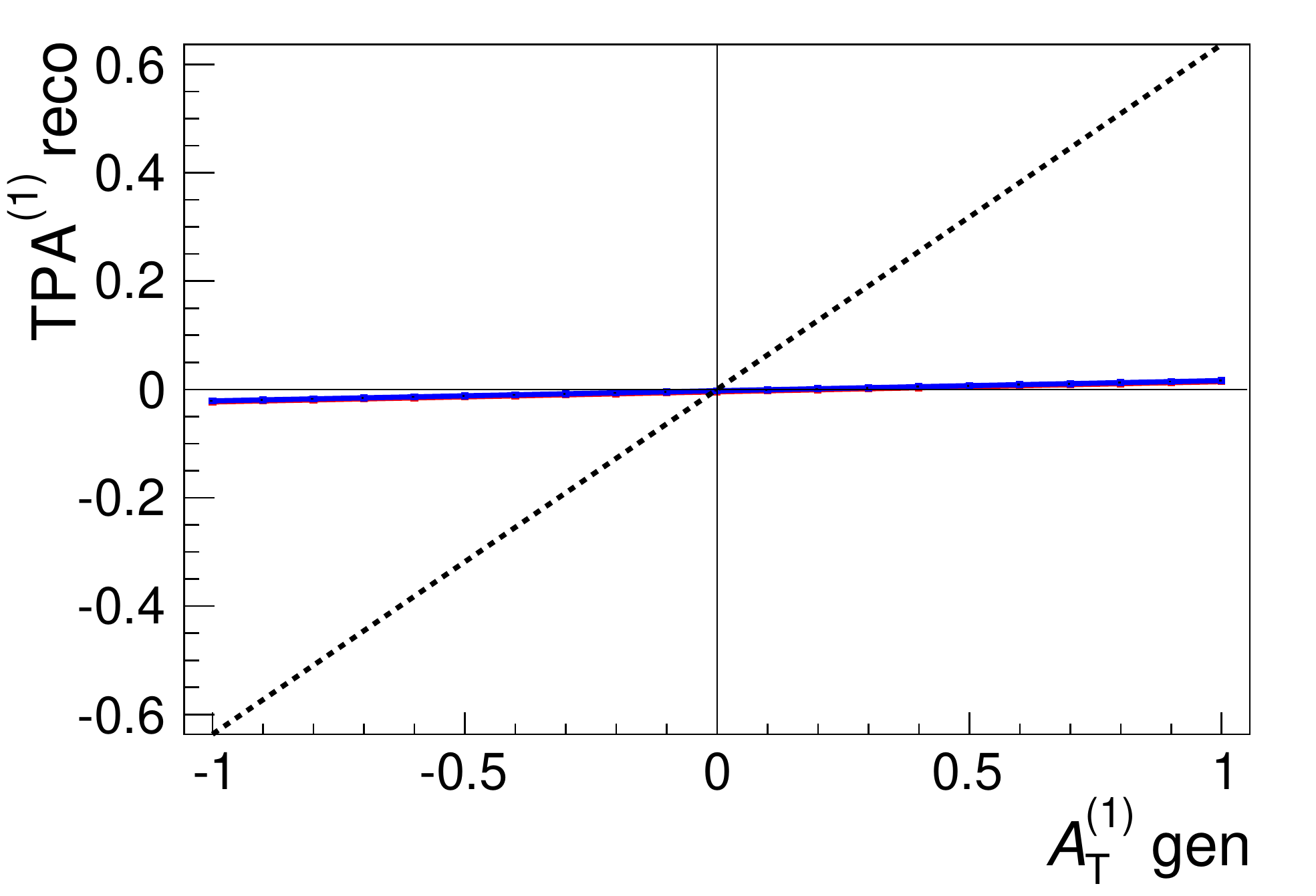}
\includegraphics[scale=0.24]{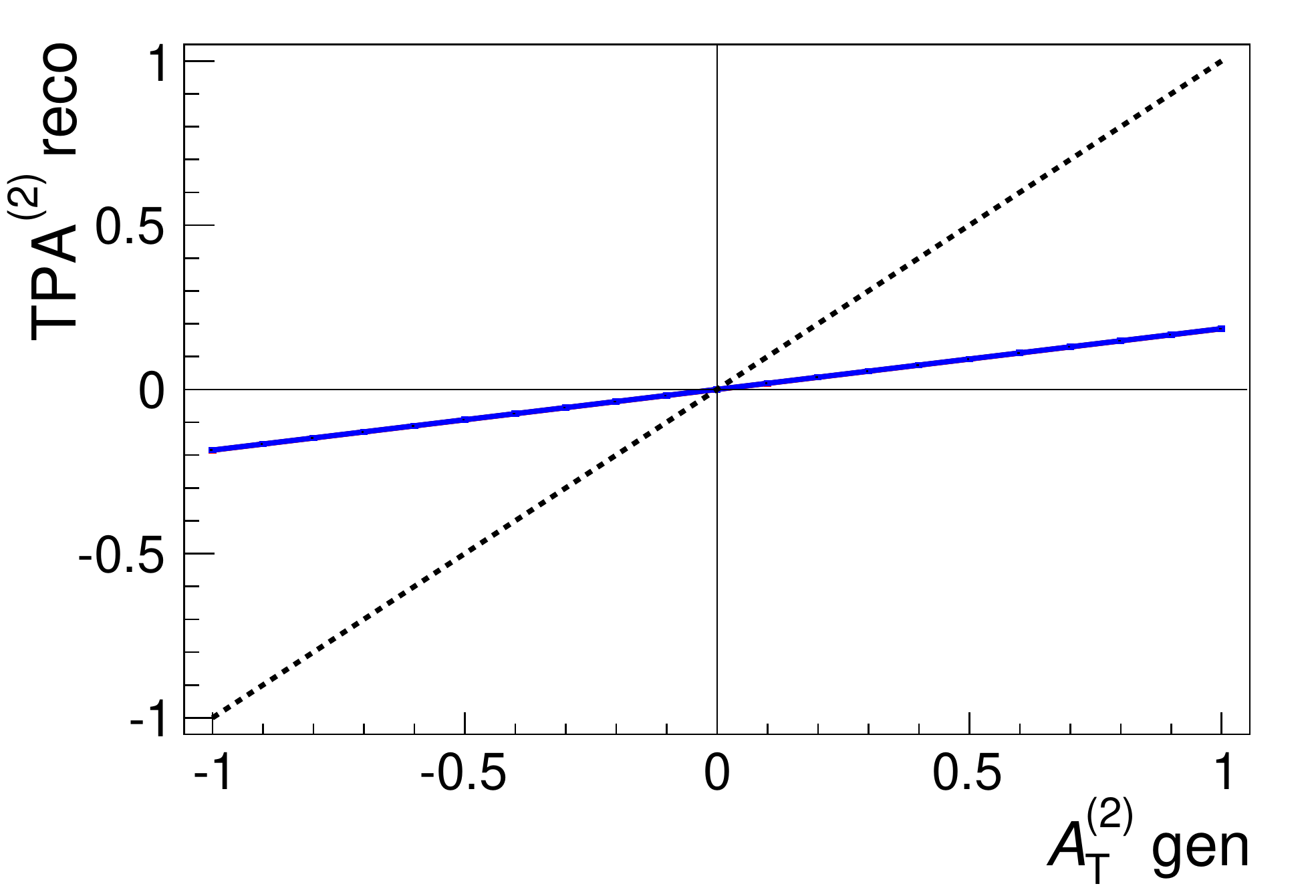}
\caption{Relation between TPAs and \CP-violating observables for (top) \BtoDstarmunu, (middle) \BtoDstartaunutopi and (bottom) \BtoDstartaunutomu decays, for three values of the corresponding \CP-conserving observables $\mathcal{A}^{(i)}_C$: (black) 0, (red) 1, (blue) -1. Lines are almost overimposed, showing the relation is independent of the $\mathcal{A}^{(i)}_C$ values.\label{fig:TPA_study}}
\end{figure}

The study is carried out for \TPA, defined in equation~\eqref{eq:TPA1}, from the distribution equation~\eqref{eq:Ac_At_distribution}, \TPAA, defined in equation~\eqref{eq:TPA2}, from the distribution equation~\eqref{eq:chi_gamma2_distribution} and \TPAz, defined in equation~\eqref{eq:TPA0}, from the distribution equation~\eqref{eq:Acz_Atz_distribution}. The $\mathrm{TPA} = f(\mathcal{A}_T)$ relations for the three \BtoDstarlnu decay modes are reported in figure~\ref{fig:TPA_study}. They are the same for different $\mathcal{A}_C$ values within uncertainties. From TPA definitions follow that for perfect reconstruction the $\kappa$ factor is $\pi/2$ for \At and \Atz, one for \Att. The decrease in precision from perfect reconstruction is summarized in table~\ref{tab:TPA_study} for the different asymmetries and decays.

\begin{table}
\centering
\begin{tabular}{lccc}
\hline
Penalty factor & \At & \Atz & \Att\\
\hline
\BtoDstarmunu & 2.8 & 2.1 & 2.3\\
\BtoDstartaunutopi & 12.3 & 4.2 & 5.2\\
\BtoDstartaunutomu & 15.3 & 4.2 & 5.2\\
\hline
\end{tabular}
\caption{Decrease in precision on the \CP-violating observables with respect to perfect decay reconstruction, as determined from the slope of the $\mathrm{TPA} = f(\mathcal{A}_T)$ relation.\label{tab:TPA_study}}
\end{table}

The decrease in precision on \At and \Atz is compatible to that obtained in the previous section using maximum-likelihood fits: the ``test'' observable \Atz can be measured in both lepton decay modes, while the huge penalty to be paid for the \At measurement in the tau decay mode prevents a useful measurement without exploiting information additional to the reconstruction algorithm. On the contrary, the small decrease in precision between \Atz and \Att shows that the effect of the integration of the $\sin2\thetaD$ terms is modest and the measurement of the \CP-violating \Att observable is viable for both muon and tau decay modes. This allows to search for New Physics \CP-violation in \BtoDstarlnu decays even at hadron collider experiments with a not prohibitive loss in sensitivity.

\section{Discussion on systematic uncertainties}
\label{sec:systematic_uncertainties}
In the proposed measurements, there are two steps which can introduce systematic uncertainties: the extraction of \BtoDstarlnu angular distributions from the template fit to the discriminating variables, via the \splot technique, and the use of simulated events for both the detector efficiency correction and the determination of the polynomial expansions.

For the first step, the use of the \splot statistical tool does not introduce additional systematic uncertainties to those related to the template fit itself, in which uncertainties in the modelling of the different contributing decays can lead to uncertainties in the fit results. LHCb \RDstar measurements~\cite{Aaij:2015yra,Aaij:2017deq} have shown that these uncertainties can be controlled down to the size of statistical uncertainties. On the contrary, the considered angular observables do not depend directly on the fit results, and it has been shown in section~\ref{sec:extraction} that they are not correlated with the discriminating variables on which the template fit is based.
Provided that this effect has to be properly evaluated, it is reasonable to expect the impact of these systematic sources to be smaller than for the \RDstar measurement.

Regarding the use of simulated events, uncertainties can follow from imprecisions in the detector simulation. The accuracy of detector simulation in particle physics experiment is routinely checked with respect to data, and remaining differences between real and simulated events are corrected exploiting suitable ``control'' decays as similar as possible to the transitions under study~\cite{Aaij:2015yra,Aaij:2017deq}. Moreover, the simulation of the detector resolution due to the reconstruction algorithms, exploited to correct the observable values, is based upon the decay kinematics (particle momenta and decay vertex position distributions), which is easy to simulate with high accuracy. No significant differences between real and simulated angular distribution are thus expected and the associated systematic uncertainties can not have a significant impact.

Summarizing, the measurement of the \Dstar polarization fraction and \CP-violation observables should not be affected by additional systematic uncertainty sources with respect to the \RDstar measurements~\cite{Aaij:2015yra,Aaij:2017deq}. Fit model and data-simulation discrepancies uncertainties, which have already been studied for the \RDstar measurements, are expected to have a smaller impact on the proposed measurements.

\section{Conclusions}
\label{sec:conclusions}
A simulation study for a forward detector geometry is performed to quantify the attainable precision on the \BtoDstarlnu angular distributions, with the use of reconstruction algorithms estimating the \Bbar meson rest frame only from information related to the detectable final-state particles. This is of particular interest for hadron collider experiments. The resolution distributions have been found to be symmetric and unbiased for \costhetaD and $\chi$ variables, which also show negligible correlations with the discriminating quantities employed for selecting \BtoDstarlnu decays, making \costhetaD and $\chi$ distributions suitable to be extracted using the \splot statistical technique.

Observables related to \costhetaD and $\chi$ variables are the \Dstar longitudinal polarization fraction \fl, the \CP-conserving quantities $\mathcal{A}^{(i)}_C$ and the \CP-violating observables $\mathcal{A}^{(i)}_T$. The latter are of particular interest being a null test of the Standard Model.

A method to correct the effect of reconstruction inaccuracies on the mentioned observables is tested on simulated \BtoDstarlnu decays. The decrease in precision due to the employed reconstruction algorithms is evaluated. According to the simulation study, the \Dstar longitudinal polarization fraction is measurable for both muon and tau lepton decay modes with a maximum penalty in sensitivity of a factor 2. This permits an additional search for New Physics in \BtoDstarlnu decays complementary to the already measured \RDstar ratio. The \CP-violating \Att observable can be measured from the associated triple-product asymmetry with a decrease in sensitivity of a factor 5, while the \At observable is measurable only for \BtoDstarmunu decays due to the form of the associated $\chi$ angle distribution.
It is also argued that systematic uncertainties associated to the proposed measurements do not have a large impact.




\addcontentsline{toc}{section}{References}
\setboolean{inbibliography}{true}
\bibliographystyle{LHCb}
\bibliography{biblio_SL}


%
%
%
%
%
%
%
%
\end{document}